# Laser-generated CuPdAgPtAu High-Entropy Alloy Nanoparticles – Thermal Segregation Threshold and Elemental Segregation

*Felix Pohl\*, Robert Stuckert, Florent Calvo, Oleg Prymak, Christoph Rehbock, Ulrich Schürmann, Stephan Barcikowski, Lorenz Kienle*

Felix Pohl, Ulrich Schürmann, Lorenz Kienle

Department of Materials Science, Synthesis and Real Structure, Faculty of Engineering, Kiel University, Kaiserstraße 2, 24143 Kiel, Germany.

E-Mail: fepo@tf.uni-kiel.de

Robert Stuckert, Christoph Rehbock, Stephan Barcikowski

Technical Chemistry I, University of Duisburg-Essen and Center for Nanointegration Duisburg-Essen (CENIDE), Universitaetsstr. 7, 45141 Essen, Germany.

Florent Calvo

Université Grenoble Alpes, CNRS, LIPHY, 140 rue de la Physique, 38402 Saint Martin d'Hères, France.

Oleg Prymak

Inorganic Chemistry and Center for Nanointegration Duisburg-Essen (CENIDE), Universitaetsstr. 7, 45141 Essen, Germany.

Ulrich Schürmann, Lorenz Kienle

Kiel Nano, Surface and Interface Science (KiNSIS), Kiel University, Christian-Albrechts-Platz 4, 24118 Kiel, Germany.

Funding: The authors thank the Deutsche Forschungsgemeinschaft (DFG) for financial support within the project Nr. 277627168.

Conflict of Interest: The authors declare no conflicts of interest.

**Abstract**

High-entropy alloy nanoparticles synthesized via laser ablation in liquid are promising for catalysis due to their ability to form simple solid solutions despite chemical complexity. In this study, noble metal HEA NPs (CuPdAgPtAu) are produced from equimolar and Cu- or Ag-enriched bulk targets. Advanced electron microscopy, XRD, and atomistic simulations are used for structural and compositional analysis. Equimolar targets and NPs exhibit a single fcc phase. In contrast, Cu- or Ag-enriched targets show phase segregation into two fcc phases, which is not observed in the synthesized NPs. Simulations predict segregation tendencies, including Ag surface enrichment and Pt core enrichment due to surface energy differences. However, experimentally, individual NPs remain compositionally homogeneous. Thermal stability studies reveal that phase segregation can be induced post-synthesis. Upon heating, Cu–Ag segregation occurs, forming a second fcc phase similar to bulk targets. These findings demonstrate that rapid quenching during laser ablation suppresses thermodynamically driven segregation and stabilizes metastable solid solutions under kinetic control. Subsequent slow heating overcomes kinetic barriers, enabling equilibrium phase formation at higher temperatures. The thermal stability of these NPs and their tunable composition, including Cu enrichment beyond equilibrium limits, make them promising for high-temperature catalytic applications while reducing noble metal usage.

# 1. Introduction

High-entropy alloys (HEAs) have been investigated as bulk materials for over 20 years due to their unique properties, such as strength, ductility, and corrosion resistance [1], [2]. More recently, they have attracted growing interest as nanoparticles (NPs), particularly for energy-related applications and catalysis, due to the interaction of multiple elements and their high surface-to-volume ratio [3], [4]. Their suitability for heterogeneous catalysis arises from the complex interplay between the different elements, yielding specific adsorption energy distribution patterns [5]. This makes HEA NPs promising candidates for nitrogen conversion reactions [6], electrocatalytic water splitting [7], and $CO_2$ reduction reactions [8], [9].

HEA NPs have been reported with a variety of compositions and different structures, for example, PtAuPdRhRu [10] or CrMnFeCoNi [11] with face-centered cubic (fcc), TiZrHfMoNb [12] with body-centered cubic (bcc), or $Ir_{0.19}Os_{0.22}Re_{0.21}Rh_{0.20}Ru_{0.19}$ [13] with hexagonal closed packed (hcp) crystal structures. Additionally, amorphous structures have been observed in ignoble metal-based HEA NPs such as CoFeLaNiPt [14]. Amorphous structures generally exhibit higher catalytic activity than crystalline structures due to their higher fraction of atoms with low coordination numbers, providing sufficient active sites and a unique electronic structure for catalysis [15]–[18].

The synthesis of HEA nanoparticles has been reported for both noble and base metals. Base metals are especially attractive due to their low cost and high abundance, while noble metals are interesting due to their higher chemical stability and selectivity. Various synthesis methods have been employed, including chemical reduction [19], solvothermal approaches [20], [21], or kinetic-controlled synthesis methods [22], such as carbothermal shock synthesis (CTS) [23], microwave heating [21], and mechanochemical synthesis [24], [25].

While CTS and microwave heating methods can produce uniformly mixed elements in HEA NPs, their restricted flexibility in generating colloidal HEA NPs limits their broader applicability, as their use is confined to the support on which they were created. Laser ablation in liquid (LAL) offers a ligand-free approach for the large-scale synthesis of colloidal elemental and alloy NPs [26], [27]. This method has been successfully applied to ignoble HEA NPs, yielding both crystalline and amorphous structures [28]–[32]. In previous work, we synthesized ignoble HEA NPs in various organic solvents, with the resulting amorphous or crystalline structure depending on the pulse duration, which served as a key parameter influencing structural formation, due to altered mechanistic processes in regard to carbon incorporation into HEA NP cores [29]. While the LAL synthesis of binary and ternary noble metal alloy NPs in organic solvents is already well established [26], and LAL synthesis of noble metal HEA NPs in water has been recently reported [33], LAL synthesis of noble metal HEA NPs in organic liquids has not been reported yet. Furthermore, compositional HEA tuning during LAL and temperature stability of the noble metal HEA have not been investigated yet.

In this study, we apply LAL to fabricate CuPdAgPtAu HEA NPs, specifically addressing the limits of solid solution formation in this multimetallic nanosystem by enrichment of both Ag and Cu, respectively, in the individual compositional systems, comparing it to the initial near-equimolar CuPdAgPtAu alloy, as has been similarly shown in binary systems like Fe-Au [34]. To minimize oxidation and exclude oxidation-driven elemental segregation, LAL synthesis was performed in purified acetone. Ag was selected for specific reasons: silver exhibits the lowest surface energy and melting point among the components [35] and should be prone to the formation of segregated phases on the particle surface in case the limit of solid solution formation in the system is exceeded. Experimental studies on LAL-synthesized AgAu NPs have also confirmed this surface segregation tendency of Ag [36]. Moreover, simulations of noble-metal HEAs and trimetallic AgAuCu clusters consistently predict surface enrichment of Ag [37]. From studies on binary and ternary alloys, it is known that Ag exhibits low solubility with elements such as Cu or Pt, which extends to Au-Pt alloys as well. For Cu-Au and Cu-Pd systems, the formation of intermetallic phases has been reported, further limiting the solubility of Ag in these alloys. [38]. Cu enrichment was selected due to its chemical properties, like the highest oxidation propensity of the elements in the Cu-Pd-Ag-Pt-Au system, but particularly due to its notable catalytic properties in copper-based catalysts, for instance, in carbon dioxide hydrogenation [39], [40]. Nanoparticle characterization includes transmission electron microscopy (TEM), energy dispersive X-ray spectroscopy (EDS), selected area electron diffraction (SAED), and X-ray diffraction (XRD) with additional *in situ* and *ex situ* heating experiments. Further Monte Carlo and molecular dynamics simulations are included, offering valuable insights into the anticipated elemental distribution and possible segregation, as well as the role of kinetics.

## 2. Results and Discussion
### 2.1 Compositional and structural characterization of HEA targets and nanoparticles

The morphology and size distribution of the NPs are shown in **Figure S1** in the Supporting Information Section S1. The particles have a spherical shape in various sizes, which were determined by analysis of TEM images. Independent of their composition (NM-Eq, NM-Cu, and NM-Ag), the particle morphologies and size distributions are similar, although the mean size increases for NM-Ag (32 ±16 nm) and NM-Cu (22 ±9 nm) compared to NM-Eq (17 ±8 nm). This change can be observed for the width of the size distribution as well, although the polydispersity index (PDI) for NM-Cu (0.16) is smaller than for NM-Eq (0.22) and NM-Ag (0.25). Structural analysis was done using XRD with Rietveld refinement (Supporting Information Section S2), and a reduced diffractogram is shown in **Figure 1**. It reveals that both the NM-Eq target and the synthesized nanoparticles exhibit a single fcc phase, indicating no significant structural segregation of the elements. The reflection profile of the target is wider than that of the NP, which is supposed to be caused by local changes in composition resulting in several small changes in lattice parameter (Supporting Information Section S2). The respective lattice constants were calculated to be $a_{target}$ = 3.96 Å ±0.02 Å and $a_{NP}$ = 3.93 Å ±0.02 Å. The small deviation may be explained by a change in the crystallite size of the NPs or lattice strain resulting from rapid heating and cooling, but overall, the average structure is maintained after ablation. Interestingly, the present results are in contradiction with earlier work by Sohn et al., in which CuPdAgPtAu bulk material synthesized by arc-melting was found to contain two fcc phases, because Pt-Cu and Pd-Cu have negative enthalpies of mixing, which favors mixing [41]. One possible explanation for this difference is a deviation from the nominal equimolar composition: less Pd and Pt can reduce the effect of negative enthalpies of mixing with Cu, resulting in a more favorable mixing behavior between Cu and Ag/Au instead. For nanoparticles synthesized via cryomilling under Ar atmosphere, a single fcc phase for the same composition has been reported [8], [42].

When examining the diffraction patterns of the Ag- and Cu-enriched bulk targets, a second phase can be observed. In both targets, these two fcc phases — distinguished by their lattice parameters — can be separated into a major and a minor phase based on their respective intensities. Neither of these two phases can be attributed to oxidation. In the case of Cu enrichment, the reflections of the major phase shift to a smaller lattice parameter of 3.78 Å ± 0.01 Å, since Cu has the smallest atomic size among these elements. Conversely, for the Ag-enriched bulk target, the opposite occurs, resulting in a lattice parameter of 4.04 Å ± 0.01 Å. Additionally, the (111) reflection splits into two distinct reflections with lattice parameters of 4.04 Å and 4.03 Å, respectively. These potentially strain-induced deviations in composition within the enriched phases affect the (111) reflection. Other reflections of the major phase show a broadening and asymmetry. This potential phase separation was resolved by Rietveld refinement into two potential phases with nearly identical lattice parameters but different crystal sizes, which may be the result of strain caused by our inherent non-uniform composition (Supporting Information S2). Noteworthily, the minor phase of the Ag-enriched target ($a_{minor}$ = 3.79 Å) matches with the major phase of the Cu-enriched target ($a_{major}$ = 3.78 Å), and vice versa, as highlighted by dashed vertical lines in Figure 1. This indicates segregation into Ag- and Cu-rich phases within the bulk material. In both cases, neither the major nor the minor phase corresponds to a single fcc phase from the equimolar composition, which can be attributed to the significant difference in lattice parameters between Ag and Cu. At equimolar composition, neither of the two elements dominates the structure, resulting in an averaged d-value with respect to NM-Ag and NM-Cu NPs major phase. The entropy of mixing is highest at this composition compared to Ag- or Cu-enriched phases. Corresponding values for the entropy of mixing, $\Delta S_{mix}$, are provided in **Table S4**. Therefore, the equimolar phase is likely stabilized due to its entropy of mixing, whereas enrichment in Ag or Cu increases the enthalpy of mixing and decreases the entropy of mixing, thus overcoming the stabilization effect at the equimolar composition [43].

While an enrichment of Ag or Cu leads to two phases in the target, the synthesized NPs are comprised of only one phase, as shown in Figure 1. For both compositions, their respective major phase forms during LAL. This difference between target and NPs can have two possible causes: Firstly, the coherently scattering domains of the minor phase could be too small and cannot be identified by XRD for the NPs. Secondly, a process or condition may occur during pulsed laser synthesis, preventing the formation of the minor phase, therefore, stabilizing the NPs into a single fcc phase. One possibility for this stabilization is the rapid heating and cooling conditions during LAL synthesis, resulting in a metastable fully mixed phase formation. During the initial stage of the plume created by fast heating with the laser pulse, lateral intermixing of both fcc phases results in a joint phase inside the plume from which they solidify. There might not be a sufficient time frame for elemental segregation [44], providing kinetic control of NP formation. Computational and experimental works with AgCu layered targets contradict this statement, as limited intermixing was found in the initial stage during the first few nanoseconds [45]. There is a limited transferability here, since in this study, a fully demixed, layered thin film target was assumed, and only the initial stage was analyzed. Our targets are neither thin film sandwiches nor fully demixed. In addition, the laser spot diameter might be larger than these phase-segregated areas, further elaborating on intermixing during the initial ablation step. Waag et al. investigated how the laser pulse duration and spot size contribute to mixing of laterally micro- and nano-segregated LAL alloy targets, with longer pulse durations and better pre-mixed ablation targets providing better mixing by the plume [46]. We used ps pulse durations, which provide less mixing compared to ns pulses, so that an immediate initial intermixing and subsequent prevention of segregation is probably mainly due to fast cooling, resulting in single-phased NPs. In contrast, the target was cooled down more slowly by convection in an argon atmosphere, which enables it to reach a thermodynamic equilibrium, resulting in the two phases observed.

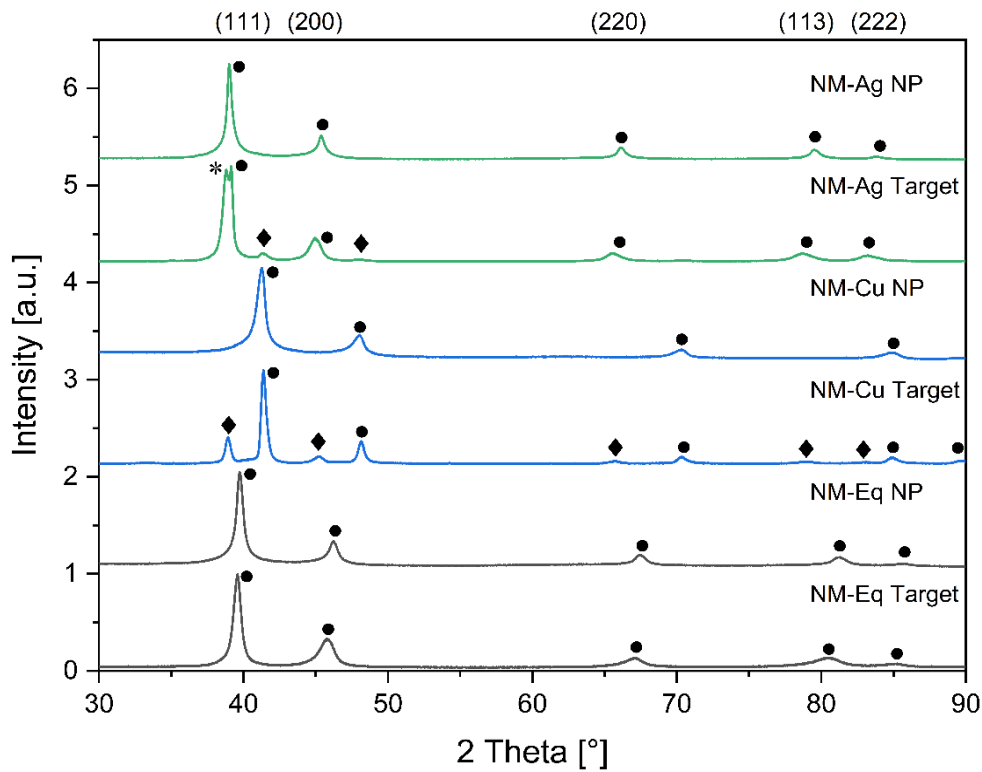

**Figure 1** X-ray diffraction patterns from targets and nanoparticles fabricated thereof by LAL. Miller indices are given above, black dots mark reflections from the major phase, diamonds mark reflections from the minor phase, and an additional reflection in the NM-Ag target is marked by an asterisk, which is included into the major phase.

To exclude the possibility of the second phase in the NP sample simply being non-detectable in XRD patterns, additional SAED measurements were performed. The corresponding diffraction patterns for NM-Eq, NM-Cu, and NM-Ag are presented in **Figure 2**, respectively. In each case, the diffraction patterns indicate the presence of a single crystalline phase. For NM-Eq and NM-Ag, the resolution may be insufficient to clearly resolve the (111)-reflections of a potential minor phase with a smaller lattice parameter from the (200)-reflections of the dominant phase. However, the absence of distinct (200)- and (220)-reflections attributable to such a minor phase supports the conclusion that only a single phase is present. In the case of NM-Cu, any minor phase with a larger lattice parameter than the major phase would be expected to produce (111)-reflections at a smaller diffraction angle. The absence of such reflections in the observed pattern further corroborates the single-phase nature of the nanoparticles.

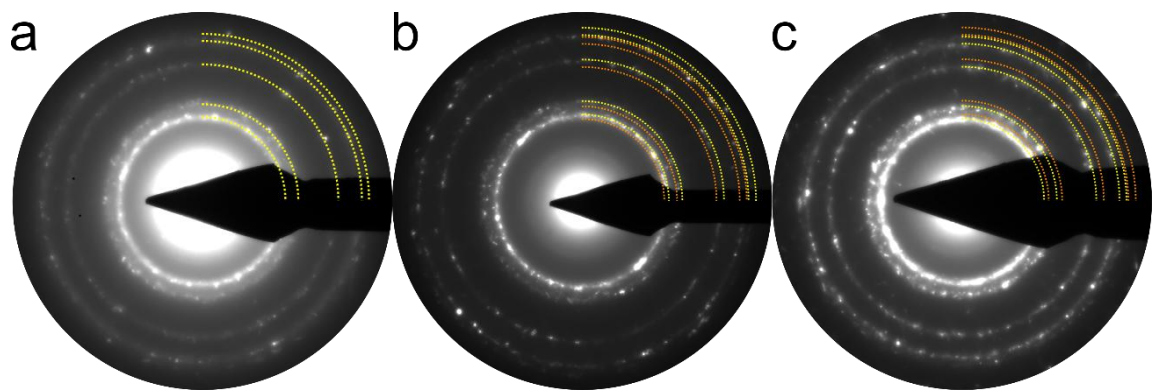

**Figure 2** SAED patterns for NM-Eq (a), NM-Cu (b), and NM-Ag (c) showing one fcc phase for each composition, dashed lines mark calculated reflection positions for each major phase (yellow) and minor phase (orange), respectively.

Furthermore, compositional analysis was done with SEM-EDS on the non-ablated target surfaces and TEM-EDS on the synthesized NPs. In **Figure 3**, the obtained compositions are given as well. The measurement for the equimolar composition shows a deviation from the nominal composition. Ag, Au, and Cu are enriched, while Pd and Pt both show some depletion. Especially, Pt is reduced to less than 10 at-%. Due to their higher melting points, Pt and Pd crystallize first during cooling. In addition, the slow cooling process enables diffusion of atoms, resulting in a thermodynamically favorable surface segregation of Ag, Au, and Cu due to their lower surface energies [35]. This is consistent with electronic structure calculations by Kristoffersen et al., who simulated AgAuCuPdPt HEA surfaces and found some Au and Ag segregation at the surface as well as Pd and Pt depletion, being one possible explanation for our observation [47]. A comparison of the overall composition using large area EDS measurements with the composition determined from the target shows a similar chemical composition, see Figure 3. Our nominal composition for each element in NM-Eq was 20 at-%. As shown in Figure 3, the contents of Pt and Pd are decreased to 9 at-% and 12 at-%, respectively, while the contents of Cu and Au increase to around 27 at-%. The Ag content approaches nominal composition with around 22 at-%. This deviation from our nominal composition can be caused by a combination of two factors: an enrichment of Ag, Cu, and Au on the surface, while Pd and Pt enrichment in the core of the target due to their different surface energies [35], [47], as well as a limited probing depth of SEM-EDS used for characterization [48]. For NM-Cu, an increase in the Au content in the target compared to the corresponding NP can be observed, while for NM-Ag, an increased Au content can be observed for the target and NP. Overall, the compositions for NM-Cu and NM-Ag deviate less from their nominal compositions than for NM-Eq.

The individual compositions of selected particles differ from the global values for NPs. Their mean value and standard deviation are shown in **Table 1** for NM-Eq, NM-Cu, and NM-Ag, respectively. The full tables are shown in the Supporting Information Section S4, **Tables S6-S8**. Outliers with a high deviation in Ag content are marked in Table S6. Since we only investigated a limited number of individual NPs, their relevance for determining the composition is not given. However, individual compositions remain of interest. The measured content of Ag in NM-Eq (Table 1) exceeds that within the target material. Additionally, some NPs show an excess of Ag, which is accompanied by a depletion of Pt (Supporting Information S4 Table S8). While the laser spot covers an area in the several tens of micrometers range, there is limited lateral mixing inside the plasma plume [46]. Therefore, lateral variance in the surface composition of the target might lead to these different NP compositions. The elemental mappings of the target surfaces for NM-Eq, NM-Cu, and NM-Ag are shown in **Figures S11-S13**, respectively. NM-Eq (Figure S11) shows some Pd segregation resulting in Pd-rich spots, while still retaining Pd mixed with the other elements. The NM-Cu (Figure S12) target surface shows small Ag-rich areas, which can be attributed to the Ag-rich minor phase determined with XRD. Interestingly, there are no regions with weaker signal intensity in the Cu elemental map, hinting towards less Cu at the same positions of the Ag-rich positions, which might be an artifact from the measurement, as the target surface is not perfectly flat. For the NM-Ag (Figure S13) target surface, Cu segregation is observed. The brighter areas in the Cu elemental map correspond with darker areas in the Ag elemental map, which confirms the segregation into an Ag-rich major phase and a Cu-rich minor phase determined by the previous XRD results. In addition, Pt-rich areas can be observed on the NM-Ag target surface, which are connected to the Cu-rich areas. The limited mixing of Ag with Pt and favorable mixing enthalpy between Cu and Pt result in the formation of the observed minor phase containing predominantly Cu and Pt [41], [49].

Furthermore, an enrichment of Ag on the targets´ surface due to its low surface energy can result in predominantly Ag ablation during synthesis with the first laser pulses, which changes with prolonged ablation. The variance of the Ag content, as shown in Table 1, hints towards a changing Ag content in different areas in the plasma plume during the ablation process, which might be caused by Ag-rich areas on the target surface. Cu and Au show a higher deviation in their elemental composition values in comparison to Pd and Pt. This hints towards a low change in the Pd and Pt contents during ablation, while the Cu and Au contents change is greater. A possible explanation might be the potential core enrichment of Pd and Pt with a corresponding surface enrichment of Ag inside the target [47], [50]. For NM-Cu, the Ag content shows the highest standard deviation again, as shown in Table 1. Overall, the standard deviation for all elements is lower compared to NM-Eq. The outliers in Table S7 show, as explained before, a lower Ag content with a simultaneous increase in the Pt content. However, these outliers cannot be seen as a hint of the minor phase being present inside the NP.

The individual composition of NPs from NM-Ag in Table 1 shows a similar behavior: the Ag content has the highest standard deviation and is, on average, higher than measured by EDS inside the target, which hints towards a diffusion of Ag to the target surface, resulting in a higher Ag content. Interestingly, some NPs show a low Ag content, which again is linked to a higher Pt content. The observed phase segregation on the target's surface (Figure S11-S13) can explain this difference in the composition. This might also be linked to the minor phase detected by XRD inside the targets. The two phases from the target might not intermix perfectly inside the plasma plume during ablation, which then transfers to Ag and Pt-rich areas inside it, from which then these Ag and Pt-rich particles form. Since we only measured a limited number of particles, this might not be a general effect and the overall composition of the particles is still close to that of the targets (see Figure 3), the individual composition is likely distributed around it.

**Table 1** Summary of individual NP composition for NM-Eq, NM-Cu, and NM-Ag with mean composition and corresponding standard deviation (sd), full data can be found in Supporting Information Section S4.

|  | Cu | Pd | Ag | Pt | Au |
| --- | --- | --- | --- | --- | --- |
| NM-Eq mean | 25.2 | 14.2 | 32.6 | 6.8 | 21.1 |
| NM-Eq sd | 7.6 | 2.5 | 13.3 | 2.1 | 6.1 |
| NM-Cu mean | 51.1 | 15.4 | 12.7 | 9.8 | 10.9 |
| NM-Cu sd | 2.5 | 1.5 | 4.6 | 2.9 | 1.8 |
| NM-Ag mean | 9.2 | 14.3 | 60.3 | 6.3 | 9.9 |
| NM-Ag sd | 2.5 | 4.9 | 12.4 | 3.5 | 4.2 |

Additional XPS measurements were performed to determine the (surface) composition of the nanoparticles. The results are shown in Figure 3. Important to note is the higher amount of Ag and Cu determined by XPS for NM-Eq in comparison to our EDS results. As XPS is a surface-sensitive technique with limited penetration depth of up to several nanometers, this suggests an Ag and Cu surface enrichment for larger particles, while particles smaller than 10 nm might contain more Ag and Cu. Since Cu is the most oxygen-affine element in our system, its higher surface content is caused by surface diffusion and subsequent oxidation, which is supported by the presence of Cu(II) peaks in XPS (Supporting Information S3, **Figure S8-S10** and **Table S5**). Ag is likely to diffuse towards the surface due to its low surface energy compared to the other elements (Pd, Pt), which diffuse towards the NP core [35], [47]. For NM-Ag, the Cu content is slightly increased compared to EDS, which can again be attributed to oxidation of Cu. The Pt content is reduced compared to EDS, which might be due to its low surface energy as well as its low miscibility with Ag [35], [49]. Therefore, Pt might be confined to the NP core. As mentioned before, Ag is likely to diffuse towards the surface, and since the nanoparticles consist mainly of Ag, its content increases. NM-Cu has a higher content of Cu compared to our EDS measurements. Again, diffusion towards the surface caused by oxidation might result in this increase. The Ag content is increased while the content of the other elements is decreased compared to EDS measurements. Ag is again diffusing towards the surface, which is not only caused by its low surface energy but also might result from the low solubility of Ag in Cu [51]. Overall, the NP composition by XPS shows an increased Ag and Cu content with a decreased Pd, Pt, and Au content compared to the target composition, which is related to differences in surface energies and oxygen affinities.

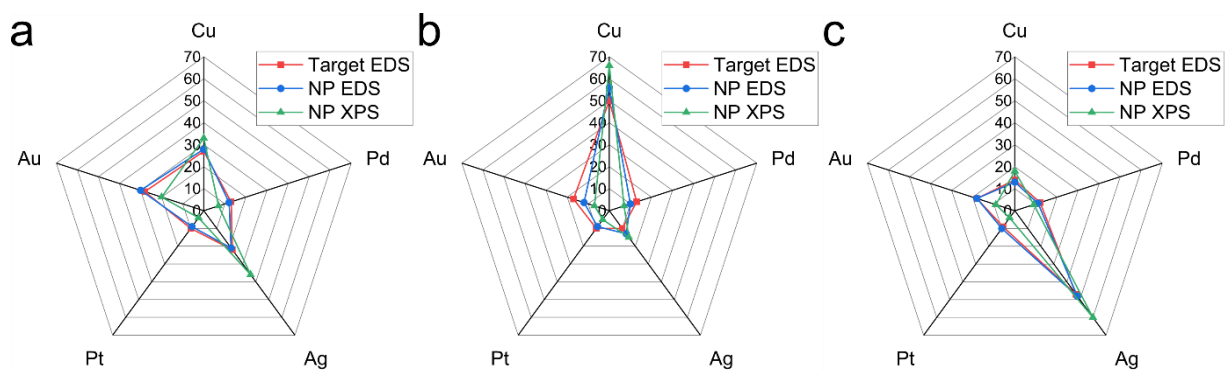

**Figure 3** Measured composition of targets and global values for the LAL-generated NPs by large area EDS and measured composition for NPs by XPS analysis for NM-Eq (a), NM-Cu (b), and NM-Ag (c).

Building on the global NP composition, intra-particulate compositional analysis has been carried out. ED measurements could not be fully conclusive in themselves to distinguish a possible phase segregation inside the NPs. Therefore, additional STEM elemental mapping experiments were conducted for all three compositions, NM-Eq, NM-Cu, and NM-Ag, as shown in **Figure 4**a-c, respectively. All five elements are distributed homogeneously inside the NPs. At first glance, this confirms a possible process stabilizing the major fcc phase. In contrast to our XPS results, no Ag surface is observed in Figure 4a, which may be attributed to a low contrast if the Ag surface layer is only 1 or 2 nm thick. Therefore, we investigated particles smaller than 10 nm by additional STEM elemental mapping experiments, since their surface-to-volume ratio is higher. The elemental map overlay in Figure 4d shows a shell around the NP that contains Ag but not Pt. This is confirmed by an EDS linescan in Figure 4e, where the intensity of Ag is increasing first, while the intensity of the other elements is not increasing yet. The thickness of this shell is between 0.5 nm and 1 nm, which was therefore not detected in elemental maps at lower magnification. The continued increase in intensity of Ag, when the intensity of the other elements increases as well, suggests that Ag may also be located within the core of the NP. Therefore, this Ag surface layer is not seen as a phase segregation, which is consistent with our XRD and ED results. The simulation of an equimolar CuPdAgPtAu NP, whose composition is similar to our NM-Eq, shows an Ag surface layer as well [52].

Since the elemental mapping and diffraction patterns of Ag- and Cu-enriched particles align with our XRD results, a kinetically stabilized structure of the particles into a single solid solution phase is likely. During synthesis, the surface of the target is transformed into an ablation plume in which all elements are mixed. This plume is confined by the surrounding liquid, which rapidly cools it down. According to Shih et al., an effective cooling rate α of $10^{11}$ to $10^{13}$ K/s is initially reached during ultra-short pulse irradiation in LAL [29], [44]. This rapid drop in temperature is known to lead to undercooling processes within the forming nanoparticles, creating defect-rich regions [53], [54], and was shown to lead to generally thermally stable nanoparticle structures, as already demonstrated for oxidic[53], [54], binary alloy [55], and also high-entropy alloy nanoparticles [29], [56] in previous studies. Additionally, the surrounding liquid can influence the elemental distribution as the used solvent can degenerate partly into different carbon species, which interact with the different elements [57]. The particle volume itself can also limit segregation, as the number of surface atoms is restricted. Consequently, we can conclude that significant segregation (even if theoretically considerable, see consideration in Supporting Information Section S3) within our studied HEA NPs is hindered by kinetic stabilization, which is one major advantage of the here utilized synthetic approach. Further, a slight increase in Ag content in HEA nanoparticle surfaces is likely to be an effect of surface energy, as previously discussed. This effect is also discussed in the next chapter in conjunction with MC and MD simulations.

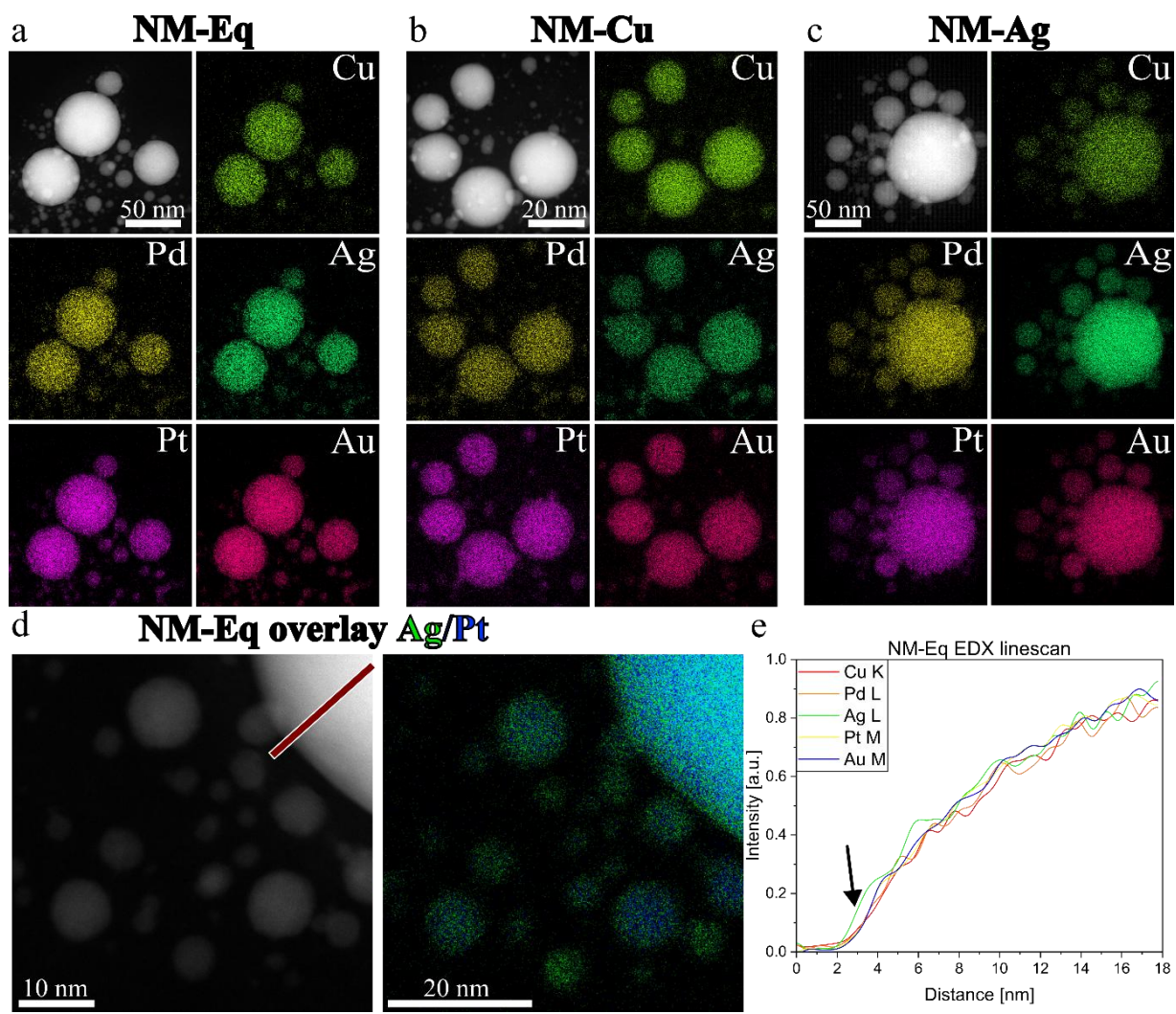

**Figure 4** EDS elemental mapping for NM-Eq (a), NM-Cu (b), and NM-Ag (c). Overlay of the EDS elemental map of Ag (green) and Pt (blue) for NM-Eq (d) with EDS linescan (e) at the marked position in the STEM-DF image (d), the arrow marks the Ag shell in the linescan.

## 2.2. Simulation results

Computational results explain how these various elements arrange within the nanoparticles under thermodynamic equilibrium conditions or after fast cooling down a gas mixture influenced by their respective mixing behavior and surface energies. **Figure 5** shows low-energy configurations obtained from the MD and MC simulations for the 4033-atom NM-Cu particles. The particles simulated by the two methods differ mostly in terms of structural ordering, the atoms remain on the fcc lattice with the MC method, while the MD simulations yield still rather amorphous nanoparticles under the relatively high cooling rate employed. Despite such important differences, both particles show clearly Ag and Au diffusion to the surface of the particles, while Pt remains in the core and exhibits a strong tendency to aggregate either in a few clusters (MC) or in one main cluster with several isolated atoms (MD). This segregation can be explained by Ag's low surface energy, while Pt has the highest surface energy of these five elements [35]. Furthermore, Ag diffusion to the surface can also be explained by the limited solubility of Cu and Ag in each other and the absence of Ag inside the NP [58]. Au diffuses towards the surface, which is caused by the low surface energy of Au, similar to Ag [35], [47]. However, Au shows a higher miscibility with the elements other than Ag. Therefore, Au can be found in deeper layers inside as well. Neither Cu nor Pd migrate to the surface like Ag or Au do, since their surface energies are higher [35]. Cu at the surface can be explained by the high Cu content inside the NP. Neither Cu nor Pd segregate from Pt or Au. Therefore, predominantly two elements remain segregating from each other: Ag and Pt.

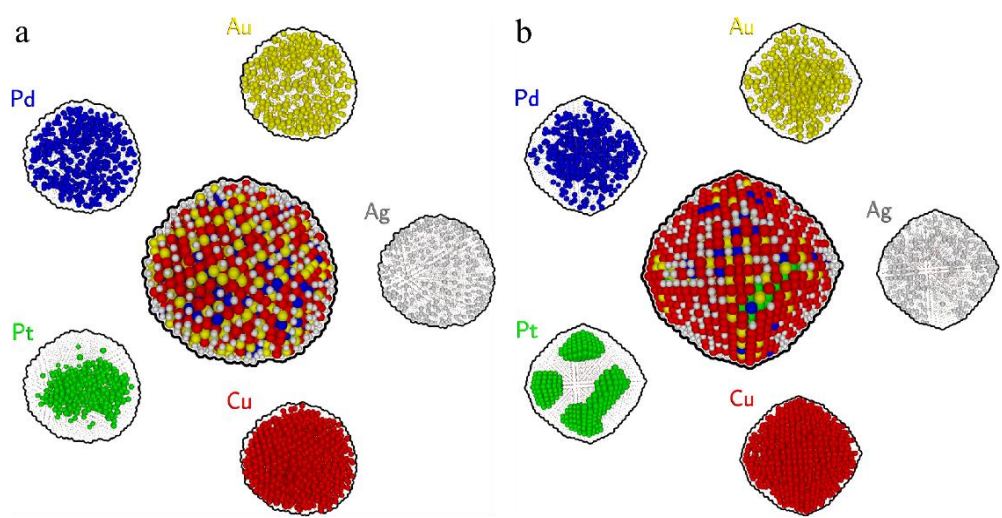

**Figure 5** Simulated particle (4033 atoms) of noble metal HEA in NM-Cu composition, (a) cooled from a random gas at 5000 K down to 300 K using Molecular Dynamics; and (b) at 300 K thermal equilibrium using Monte Carlo on a finite (Wulff shape) face-centered cubic lattice.

In **Figure 6**, the corresponding simulated particles from MC and MD are shown for the NM-Ag particle. The qualitative picture is similar to that obtained for the NM-Cu system, with Ag diffusing to the surface of the particles due to its low surface energy, while Pt clusters into the core into a few fragments. Interestingly, Ag seems to segregate from the other elements as well, resulting in a AuCuPdPt cluster inside the particles, which can be seen for the thermal equilibrium from MC simulations. In the out-of-equilibrium MD simulations, the segregation of silver towards the particle surface may not be as strong as under equilibrium conditions, presumably due to greater mixing with gold and palladium inside the NP and subsequent trapping during the early stages of particle formation. Segregation of Ag from Cu and Pt is expected due to their limited miscibility, but Au/Pd and Ag are completely miscible.

The propensities of the various elements to distribute into few or many fragments were quantified from the statistical percolation analysis [52], leading to the distributions of the numbers of connected fragments shown in **Figure 7** for the NM-Eq, NM-Cu, and NM-Ag particles. In the MD case, the fragment analysis only included the final part of the trajectory for which the particle was entirely condensed, at T<1000 K. Generally, the number of fragments of a given element is low when the atoms of this element are agglomerating to clusters. Therefore, a high number of fragments is associated with a broader distribution of the given element inside the particle. For all three compositions, Pt has the lowest number of fragments, which describes the segregation of Pt and the formation of sub-clusters, here located in the core of the particles. At equimolar composition, the number of fragments for Ag is also low compared to Au, Cu, and Pd due to the described segregation of Ag to the surface [50], [52]. Comparison with the ideal, noninteracting limit further indicates that copper is the element behaving closest to a true solid solution, with platinum and silver lying farthest away from it. Upon enrichment by Cu or Ag, the respective numbers of fragments of the same element will decrease as well, since this element forms the majority of the given atoms. This explains the trends found in Figure 7(b) and (c), in which the numbers of fragments for Ag in NM-Ag and Cu in NM-Cu are significantly lower than at equimolar composition. For these systems, the few (but large) fragments are also the prediction in the solid solution limit. For NM-Eq and NM-Ag systems, Cu has the highest number of fragments, which hints towards little segregation of Cu from the other elements. This result is consistent with the similarity with perfect solid solution behavior predicted for this element at equilibrium, which is interesting in itself since Cu has a low solubility in Ag and is known to form superstructures with Pd, Pt, and Au [38]. The simulations also provide estimates of the mixing entropy based on the statistics of nearest-neighbor pair distributions [52]. The results, given in Table S4 of the Supplementary Material, generally agree with the experimentally determined values, although the slightly lower values in the NM-Ag case indicate that phase separation is exaggerated in the simulations. The entropy of mixing, as well as a negative enthalpy of mixing between Cu and Pd, can enable the high solubility of Cu in Pd, Pt, and Au elements [41]. At this stage, the atomistic simulations confirm the general trends expected from the known properties of the alloys and surface energies, namely an enhanced segregation of Ag to the surface of the particles and a significant segregation of Pt towards the core especially near equimolar composition, while enriching the particles with either Ag or Cu should result into an even strong segregation into a Pt-rich core and an Ag- or Ag/Cu-rich surface depending on the composition.

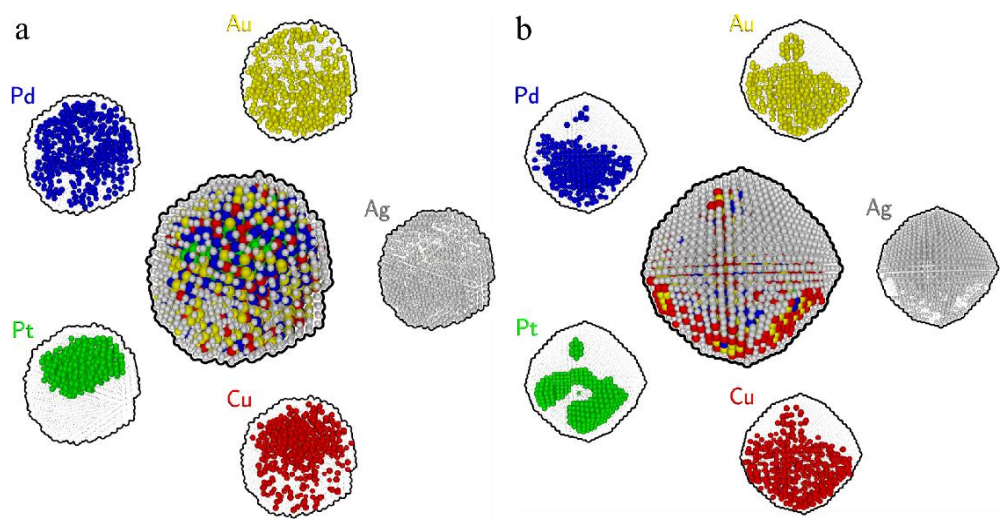

**Figure 6** Simulated particle (4033 atoms) of noble metal HEA in NM-Ag composition, (a) cooled from a random gas at 5000 K down to 300 K using Molecular Dynamics; and (b) at 300 K thermal equilibrium using Monte Carlo on a finite (Wulff shape) face-centered cubic lattice.

Elemental mapping results and EDS linescan could already prove a surface segregation of Ag in NM-Eq. Our simulation results can further emphasize this for NM-Cu and NM-Ag, which was only hinted at by XPS results. While elemental mapping did not show Pt segregation towards the core of our NPs, it is likely to be seen by comparing the XPS and EDS results in Figure 3. Contrary to our simulation results, the elemental mapping in Figure 4 shows an overall good miscibility of Pt inside our NPs. This might be caused by an interplay of different factors, which are not included in our simulations: during synthesis, the NPs are surrounded by organic liquid, which can deteriorate into different carbonyl species [57], which interact with our NPs, influencing the surface energy-driven segregation. Oxidation-driven Cu diffusion towards the surface influences the distribution of other Cu inside our NPs. Additionally, there are deviations from the ideal composition assumed in the simulations and small localized compositional deviations in the plasma plume, resulting in NPs with individual different compositions. Including the results from the experiment, and in particular accounting for the possible role of chemistry and the solvent, are needed for improving future simulation results. Furthermore, Pt has the highest melting point of the five elements, which results in a higher activation energy for diffusion. Our NPs cool down very fast, which can stop Pt diffusion earlier than assumed [59]. Comparing experimental results with our simulation results of thermal equilibrium (MC) and non-thermal equilibrium (MD), it is clear that our NPs are kinetically stabilized into a non-equilibrium state during synthesis.

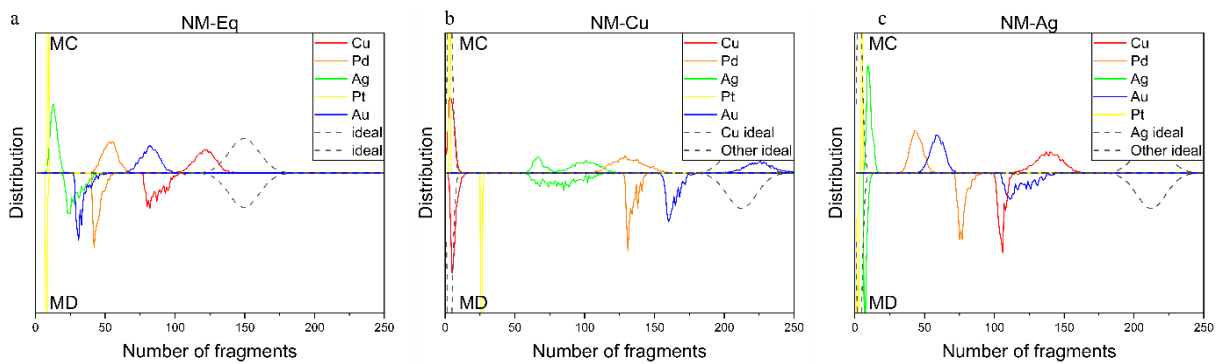

**Figure 7** Distributions in the numbers of fragments for all elemental clusters, as predicted from MC and MD simulations (a) at equimolar composition; (b) in the Ag-enriched particles; (c) in the Cu-enriched particles. The dashed lines are the distributions predicted for the ideal, noninteracting solid solution particles.

## 2.3 Thermal stability of nanoparticles

As discussed above, the synthesized particles, including the non-equimolar variants, show no sign of segregation except for a thin Ag layer on their surface. In addition, our simulations predict Ag segregation towards their surface and Pt segregation into their core, which is not supported by our experimental results. This suggests a kinetic stabilization of the NP. To further substantiate the kinetic stabilization proposed above, which may impede Ag/Pt segregation as observed in our simulations, heating experiments were carried out. Since we do not know if and at which temperature segregation occurs, first, an *in situ* TEM heating experiment was performed. Cu-enriched NPs were selected for *in situ* heating, since the lattice parameters for its minor phase are larger, therefore being more visible in the diffraction pattern. Electron diffraction patterns at selected temperatures are shown in **Figure S14**: at 200 °C, up to 400 °C, no changes can be detected. The marked reflection rings can be attributed to the major fcc phase of NM-Cu with a lattice constant $a_{NP, major}$ = 3.80 Å. At 430 °C, an additional reflection appears, which we attribute to the minor phase. By heating further, this reflection gets more defined at 495 °C. Knowing a segregation temperature from our *in situ* heating of NM-Cu NPs, additional *ex situ* experiments for all three compositions were performed at 550 °C, lying above the determined segregation temperature. The corresponding electron diffraction patterns before (top half) heating and after (bottom half) heating are shown in **Figure 8**a-c for NM-Eq, NM-Cu, and NM-Ag, respectively. Only for Cu-enriched NPs is a difference in the diffraction patterns directly visible. This might be attributed to the low spacing between reflection of the segregating phases inside the other two samples. In Figure 8b, additional reflections next to the (111)-reflection ring of the major phase can be seen after heating, hinting at the presence of an additional phase. The lattice constant obtained from this reflection ring is $a_{NP, minor}$ = 4.10 Å, while the major phase has a lattice parameter of $a_{minor}$ = 4.00 Å. The lattice parameter obtained by XRD from the minor phase in the NM-Cu target is $a_{target, minor}$ = 4.02 Å. The measured $a_{NP, minor}$ value can be attributed to a pure Ag cluster (a ≈ 4.10 Å [60]), which hints towards an Ag phase segregation forming. By analyzing the (220)-reflections for NM-Eq and NM-Ag, we can determine a phase segregation in these samples as well. The lattice constant for the major phase and the minor phase in NM-Ag are $a_{NP, major}$ = 4.05 Å ±0.01 Å and $a_{NP, minor}$ = 3.80 Å ±0.02 Å, respectively. Surprisingly, NM-Eq segregates into two phases as well. The lattice parameters of these two phases are $a_{NP, 1}$ = 4.04 Å ±0.02 Å and $a_{NP, 2}$ = 3.82 Å ±0.02 Å, which are different from the lattice of the original one-phase structure $a_{NP,Eq}$ = 3.92 Å ±0.02 Å. Comparing these two lattice parameters with the ones of the major ($a$ = 3.78 Å ±0.01 Å) and minor ($a$ = 4.02 Å ±0.01 Å) phases from the NM-Cu target hints towards a segregation into Cu- and Ag-rich phases. No segregation occurred in our target, which is assumed to be in thermodynamic equilibrium for the bulk. Therefore, an additional factor needs to be included for the NM-Eq NPs, resulting in this elemental segregation after heating. NPs have a higher surface-to-volume ratio than bulk materials, and due to their small size, their melting point is at lower temperatures [61], [62]. Both factors are advantages for a higher diffusion rate in comparison to the bulk. Additionally, diffusion of atoms results in a greater change at the nanoscale compared to large scale bulk targets. This small-scale segregation could occur in the bulk as well, where it is simply not detected. Due to the high surface-to-volume ratio, the surface energy has a stronger influence on the free energy for our NPs, which favors Ag with the lowest one and Pt with the highest one [35], resulting in a driving force for Ag diffusion towards the surface and Pt diffusion to the core. In addition, the low miscibility of Ag with Cu [58], [63] and the negative enthalpy of mixing for Cu-Pt, favoring Cu-Pt mixing [41], create an additional driving force for elemental segregation between Ag and Cu elements, which then results in these two phases, an Ag-rich and a Cu-rich phase.

For all three samples, their respective elemental mapping results after heating are shown in Figure 8d-f. By looking further, the two elements dominating this segregation are Ag and Cu. In NM-Cu and NM-Ag, differences in the elemental map of Pt can be observed, which overlap with Cu-rich regions in the Cu elemental maps. This is a further hint towards favorable Cu-Pt mixing. This effect is not observed for NM-Eq.

The segregation upon heating emphasizes the likelihood of kinetic stabilization of the NP structure during pulsed laser synthesis and confirms our earlier analysis. Ag and Cu also have limited mutual solubility, and their binary alloys show spinodal decomposition. Therefore, it is plausible that these two elements in particular show segregation [63]. While we observe Pt segregation to a small degree, it is below the stronger Pt segregation predicted by our simulation results, in which Pt segregation towards the core of the particles and Ag surface segregation were predicted to occur from both the MC and MD approaches, although on significantly smaller nanoparticles. One possible explanation is the small size of Cu atoms, which enables better mobility compared to Pt. Therefore, Cu would diffuse first, and Pt segregation will only occur at higher temperatures. *In situ* heating experiments of AgCu alloys by Ummethala et al. [63] showed spinodal decomposition at temperatures of 180 °C, which is probably not high enough for Pt diffusion as suggested by simulations from Front et al. [64]. Also, inaccuracies of our underlying atomistic many-body potential can never be ruled out, the model from Zhou et al. [65] not being particularly fitted to reproduce the properties of finite-size noble metal nanoparticles [52].

The high temperature stability of the copper-enriched nanoparticle system, where more than 50% of all atoms are copper (stoichiometrically $Cu_x(PdAgPtAu)_y$ NPs, with $x > y$), may be relevant for high-temperature applications of such single-phase HEA NPs, in particular if the reaction benefits from copper enrichment. Recently, the Ludwig group explored nanostructured high-entropy alloys in four different key electrochemical reactions, including the solid solution system Cu–Pd–Pt–Ru, documenting the effect of copper enrichment [18]. Although temperature stability and stability under electrochemical potential do not structurally correlate, there are indications that HEAs provide higher durability in electrocatalysis [66]. Moreover, basic studies on Pt alloy catalysts concluded that the temperature is a critical parameter governing the stability of the Pt-alloy electrocatalysts, linked to the ability of the nanoparticles (NPs) to retain the less noble metal [67].

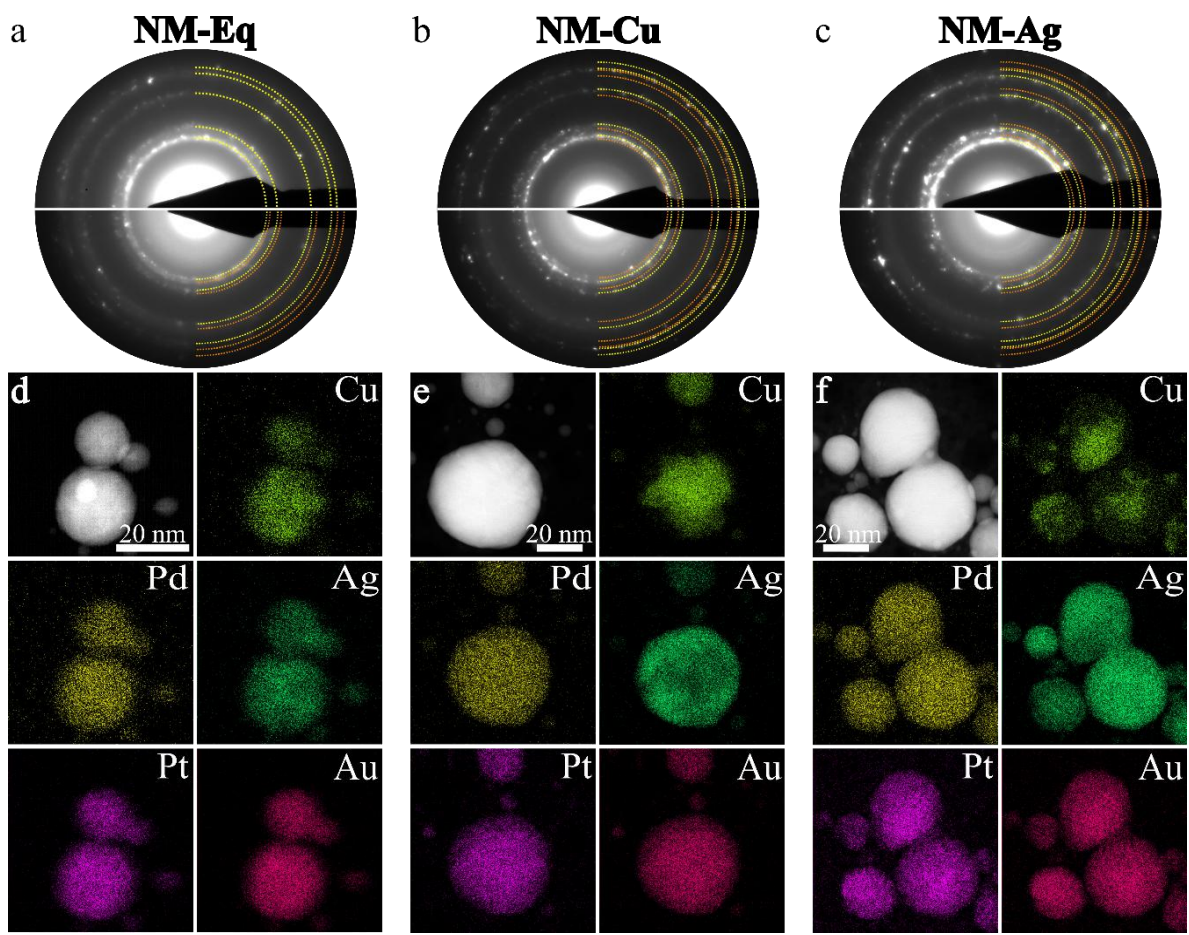

**Figure 8** SAED pattern before (top half) and after heating (bottom half) for NM-Eq (a), NM-Cu (b), and NM-Ag (c), dashed lines mark calculated reflection positions for each major phase (yellow) and minor phase (orange), respectively. Elemental map of NM-Eq (d), NM-Cu(e), and NM-Ag (f) after heating.

## 3. Conclusion

Kinetically controlled nanoparticle synthesis gives access to single-phase alloy nanoparticles. Although pulsed laser ablation of noble metal HEA targets in liquid belongs to that synthesis class, with multiple examples of HEA NP fabrication, it has not been studied yet to what extent biphasic noble metal HEA targets or significant enrichment of one element in the quinary HEA system affects NP composition and structure. Moreover, assessing the temperature stability of the HEA NPs is relevant not only to assess the metastability of the formed NPs, but also for application relevance in catalysis.

We studied noble metal HEA nanoparticles synthesized by LAL using acetone as solvent with three distinct compositions: nearly equimolar, Cu-enriched, and Ag-enriched, with the enriched compositions having about 50% of all atoms in the otherwise equimolar alloy (approx. $Cu_6(PdAgPtAu)_4$ and $Ag_5(CuPdPtAu)_5$, respectively). This enrichment of Ag and Cu resulted in a segregation into two fcc phases inside the target material, which was not observed for the equimolar composition or for the synthesized NPs. *In situ* heating experiments validated the kinetic stabilization of the NPs into a single solid-solution phase during their formation, despite the enrichment of individual elements, like Cu and Ag, and the consequent lowering of mixing entropy. Although the average elemental composition of the nanoparticles reflects that of the ablation target, the composition of individual particles varies, which can be attributed to local compositional inhomogeneities within the target material. In the bulk target, enrichment of Ag and Cu reaches the thermodynamic miscibility limits, resulting in phase separation into a two-phase microstructure with an Ag-rich fcc and a Cu-rich fcc phase. In contrast, such phase segregation is absent in the laser-synthesized nanoparticles, indicating that their structure is governed by kinetic stabilization rather than thermodynamic equilibrium. This kinetic stabilization can be exploited to develop catalysts with pronounced durability, while simultaneously maximizing the Cu content relative to the other noble metals, thereby substantially reducing material costs and enhancing sustainability.

*In situ* and ex *situ* heating experiments demonstrate that upon thermal treatment, the kinetically stabilized nanoparticles undergo phase separation into two distinct phases, namely an Ag-rich and a Cu-rich phase, analogous to those observed in the bulk material. The segregation observed in the initially equimolar nanoparticle system further highlights the metastable nature of the kinetically stabilized structure. This behavior is particularly relevant for catalytic applications, as it shows stability of the particles at typical temperatures for thermal catalysis, for e.g., $CO_2$ reduction. Important to note, the segregation temperature of 500 °C is above a typical application temperature of 200 °C, but segregation occurred for our *ex situ* experiments at 550 °C after a total time of 40 min. Therefore, segregation can occur at lower temperatures on a longer time scale, which should be elaborated in future experiments.

Surface enrichment of Ag, as revealed by EDS elemental mapping and XPS, is consistent with the trends predicted by atomistic simulations, suggesting that such simulations may be employed to predict surface compositions and guide the identification of promising catalytic formulations. While simulations predict Pt diffusion toward the nanoparticle core, this behavior could not be conclusively confirmed by elemental mapping; however, XPS measurements provide indirect evidence supporting this tendency. Future focused ion beam cuts of NPs could help in resolving these surface and core enrichment of certain elements. For the remaining elements, the experimentally observed behavior deviates more strongly from the simulations, despite overall indicators such as the mixing entropy to be correctly described. This indicates possible limitations of the employed interatomic potentials, as well as the influence of size-dependent effects, or the liquid environment, that are not captured by the model. Extended heating experiments could provide the activation energy for phase transitioning, which can be used for improving future simulations. Additionally, time-resolved heating experiments will give insight into the phase transition kinetics, offering valuable information about potential driving factors for segregation.

## 4. Experimental

*Fabrication of HEA bulk and nanoparticles*

High-entropy alloy bulk targets with the nominal compositions $Cu_{20}Pd_{20}Ag_{20}Pt_{20}Au_{20}$ (NM-Eq), $Cu_{50}Pd_{12.5}Ag_{12.5}Pt_{12.5}Au_{12.5}$ (NM-Cu), and $Cu_{12.5}Pd_{12.5}Ag_{50}Pt_{12.5}Au_{12.5}$ (NM-Ag) were produced as feedstock material for LAL. The stoichiometric targets used for each HEA NP synthesis were produced by weighing and heat-treating metal granules of Cu, Pd, Ag, Pt, and Au (Evochem, purity 99.95–99.99%) inside an arc-melting furnace for melting and sintering. To avoid oxidation, this whole process was kept under an argon atmosphere. After sintering, the targets were remolten three times, ensuring homogeneity and uniform phase formation.

All HEA NPs were synthesized in acetone (VWR, purity ≥ 99.8%), which was purified by dewatering (molecular sieve 4 Å, Carl Roth), distillation, followed by degassing with argon to minimize contamination and oxidation of the HEA NPs. Nanoparticle colloid fabrication was performed inside a self-designed stirred batch reactor with a volume of 30 ml, with a laser ablation time of 10 min, yielding colloids with mass concentrations between 100-200 mg L$^{-1}$. A Nd:YAG laser (Ekspla, Atlantic Series, 10 ps, 1064 nm, 100 kHz, 0.15 mJ, 0.1 Jcm$^{-2}$) was chosen for synthesis. The laser beam was moved on the target with a galvanometric scanner (100 mm focal length) in a spiral pattern, while lateral inter-pulse distances were set to avoid interactions between the cavitation bubbles and consecutive pulses. A power meter (PowerMax PM30, Coherent) was used to measure the average laser power behind all optics. Differential weighing of HEA bulk targets was applied for the determination of ablated mass and resulting colloid concentration.

*Material characterization*

TEM analysis, including selected area electron diffraction (SAED) and high-resolution TEM (HRTEM) of the generated HEA NPs, was conducted using a Tecnai F30 STwin G$^2$ (300 kV acceleration voltage) equipped with a Si(Li) detector (EDAX system). Chemical analysis, featuring elemental mapping and line scanning, was conducted at a probe-corrected JEOL JEM-ARM200F NEOARM scanning transmission electron microscope operated at 200 kV (cold-FEG) equipped with an energy-dispersive X-ray spectroscopy (EDS) system containing a dual silicon drift detector system. *In situ* heating experiments were also performed at the same microscope using a Lightning HB+JEOL holder from DENSsolutions. For TEM analysis, all samples were prepared by drop-casting the NP colloid on silicon nitride films (TED Pella Inc., 35 nm, 70x70 μm aperture). *In situ* heating experiments were performed using a wildfire Nano-Chip GT from DENSsolutions with a silicon nitride film as a substrate. Following the drop-casting, all samples were dried in ambient air for 1 min using an infrared lamp (Philips Infrared PAP38E, 150 W) and subsequently stored under low vacuum to avoid further contamination and oxidation. Additional *ex situ* heating experiments were performed using a self-constructed vacuum furnace. All three samples were mounted simultaneously inside the vacuum furnace and heated up to 550 °C for 30 min with subsequent cooling to room temperature in 10 min under a pressure of $2\times10^{-6}$ mbar.

Target characterization was conducted by X-ray powder diffraction (XRD, Bruker D8 Advance, Cu Kα with λ = 1.54 Å) in reflection mode in a 2θ range of 5 to 130° with a step size of 0.01° and a counting time of 1.2 s, scanning electron microscopy (SEM) (Thermo Fischer, Scientific Apreo S, LOVac, operating at 10 kV), including energy dispersive X-ray spectroscopy (EDS) (resolution <129 eV, detector area 100 mm$^2$) analysis to confirm global composition, elemental distribution, and crystal structure. HEA NPs were characterized via XRD by drying drop casting, whereby concentrated colloidal HEA NPs with comparable masses were placed on a single-crystal Si sample holder to minimize scattering. The measurements were performed with the same diffractometer (Bruker D8 Advance) in a 2θ range of 20 to 90° with a step size of 0.02° and a dwell time of 8 s. For the qualitative phase analysis, the Bruker software Diffrac

Suite EVA V7.1 was used. To calculate the lattice parameters and the average crystallite sizes, a quantitative Rietveld refinement was performed with the Bruker software TOPAS 7.0, after the instrumental characterization with a microcrystalline powder LaB6 (SRM 660b of NIST, a = 4.15689 Å) was achieved.

X-ray photoelectron spectroscopy (XPS) measurements were conducted with a VersaProbe IITM from Ulvac-Phi using the Al-Kα line at 1486.6 eV and a spot size of 100 μm with an energetic resolution of 0.5 eV. A dual-beam charge neutralization and a hemispherical analyzer (at an angle of 45° between sample and analyzer) were used for the measurements. All high-resolution spectra were corrected according to the binding energy of graphitic carbon (284.8 eV), determined during deconvolution of the C 1s spectrum of each sample. Peak deconvolution was performed using the CasaXPS software, applying a Shirley-type background. Quantification of relative compositional values of all metals was done by analyzing the 2p (Cu), 3d (Ag, Pd), and 4f (Au, Pt) spectra.

*Atomistic simulations*

In addition to our experimental results, atomistic simulations of model NPs were conducted to shed light on the preferred structures produced using a realistic interaction model and the possible role of kinetics on these predicted structures. Two complementary computational strategies were employed, both assuming fixed numbers of atoms and relative compositions in the Cu-Pd-Ag-Pt-Au elements enriched at 50% in silver (NM-Ag) or copper (NM-Cu), leaving the four remaining elements at the same relative amount of 12.5% each, with additional simulations at equimolar composition (20% for each element) being conducted for comparison. A total size of 4033 atoms was chosen as this value matches the number needed to exactly fill a 6-shell truncated octahedron based on the fcc lattice, which is the expected Wulff shape.

Monte Carlo (MC) simulations with atom swap moves were used to sample chemical equilibrium in the NPs at room temperature, assuming the atoms initially lie exactly on this Wulff lattice structure. Conventional atomic moves were attempted with 90% probability, atom swap moves between random atoms of different metals with the remaining 10% probability, moves being accepted according to the conventional Metropolis criterion. For each of the three systems studied, the MC simulations consisted of $2 \times 10^5$ cycles, with one cycle equivalent to 4033 individual moves. Averages were accumulated after $5 \times 10^4$ MC cycles. Besides low-energy structures, a statistical analysis of connected fragments [52] was also performed as a way to quantify the propensity of the various elements to aggregate together, or conversely distribute in a diluted way into the material. Fragments are here defined based on nearest-neighbor connectivity, two atoms being considered as nearest neighbors if their distance is shorter than 3 Å.

Alternatively to equilibrium MC simulations, unbiased molecular dynamics (MD) simulations were also undertaken, assuming, as in LAL conditions, that the metallic atoms are initially vaporized before cooling down to form the nanoparticles. Here, the initially hot gases (5000 K) with the same prescribed elemental compositions as in the MC simulation were linearly cooled down to 300 K at the fast cooling rate of $10^{11}$ K/s, which is quite higher than the experimental value, while sufficiently low for the particles to achieve reasonable crystallization, although not low enough to reach the Wulff crystal structures.

Both MC and MD simulations employed the many-body embedded-atom interaction model of Zhou and coworkers [65], which has been found to be rather successful in simulating mixtures of the present noble metals [50].

Finally, additional MC simulations were also performed in the simplified case where interactions are neglected altogether, keeping a lattice model for the atomic positions. The statistical distributions of fragments obtained in this approximation provide reference data in the limiting case of perfect solid solution NPs.

**Acknowledgements**

The authors gratefully thank Dr. Kateryna Loza for her experimental and analytical support with our SEM measurements and Florian de Kock for the preparation of the bulk material used in this work. XPS measurements and support by the Interdisciplinary Center for Analytic on the Nanoscale (ICAN) of the University of Duisburg-Essen (DFG RIsources reference: RI_00313), a DFG-registered core facility (Project Nos. 233512597 and 324659309), is gratefully acknowledged.

**Data Availability Statement**

The data that support the findings of this study are available from the corresponding author upon reasonable request.

**Supporting Information**

Supporting Information is available from the Wiley Online Library or from the author.

Noble metal ($Cu_xPd_yAg_zPt_yAu_y$) high-entropy alloy nanoparticles synthesized via laser ablation in liquid are characterized by advanced (S)TEM, XRD, and atomistic simulations. Rapid quenching during laser ablation suppresses thermodynamically driven segregation and stabilizes metastable solid solutions under kinetic control. Thermal stability studies reveal that phase segregation can be induced post-synthesis, overcoming kinetic barriers to enable equilibrium phase formation at higher temperatures.

**Laser-generated CuPdAgPtAu High-Entropy Alloy Nanoparticles – Thermal Segregation Threshold and Elemental Segregation**

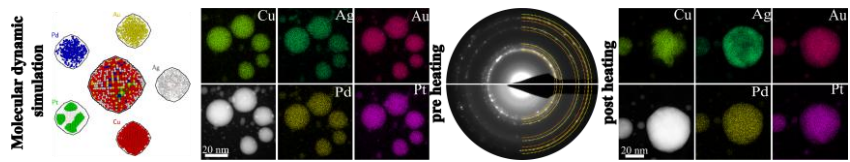

**Supporting Information: Laser-generated CuPdAgPtAu High-Entropy Alloy Nanoparticles – Thermal Segregation Threshold and Elemental Segregation**

*Felix Pohl\*, Robert Stuckert, Florent Calvo, Oleg Prymak, Christoph Rehbock, Ulrich Schürmann, Stephan Barcikowski, Lorenz Kienle*

**Section S1: TEM overview images and size distribution**

In **Figure S1**, TEM overview images of each sample and the corresponding size distribution are shown. For all samples, the majority of particles are less than 60 nm in diameter, while a few particles have diameters greater than 100 nm. The mean size of the particles increases with enrichment in Cu and Ag.

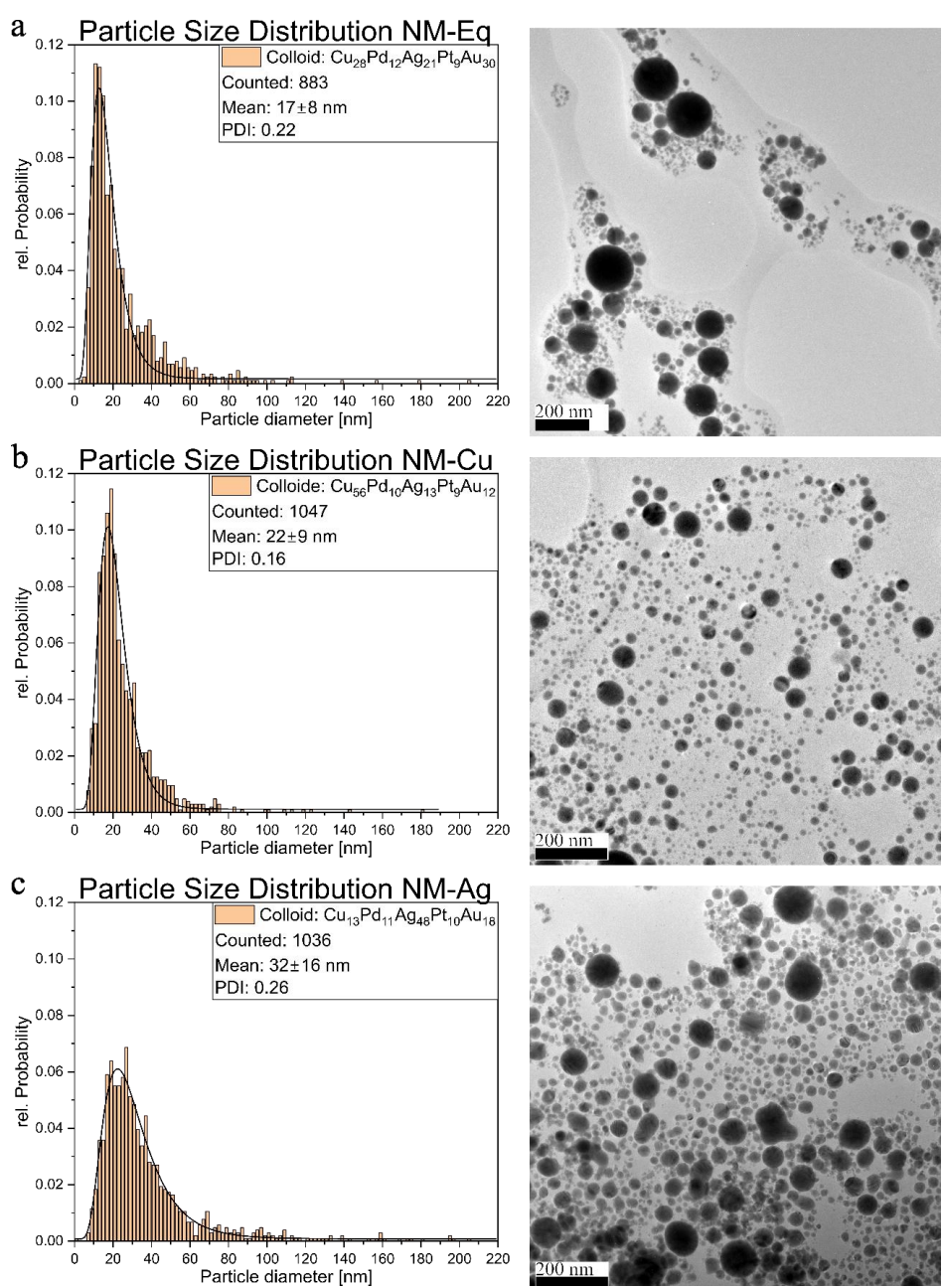

**Figure S9** Overview images and size distribution of NM-Eq (a), NM-Cu (b), and NM-Ag (c).

## Section S2: Rietveld Refinement of XRD for Targets and NPs

Rietveld refinement of our XRD data for the targets and NPs was performed to determine the corresponding lattice parameters for each sample. Lattice distortion is present in both targets and NPs because the five elements have different lattice parameters, although they share a fcc lattice structure. Small local variations in composition further influence the lattice parameter, leading to an asymmetric shape of the measured reflections in XRD. Therefore, achieving highly accurate refinements that fully describe the structure is challenging. As a result, several phases had to be refined within our structures to minimize the R factor. We summarized them into either one phase (for NM-Eq target and all NPs) or two phases (for NM-Cu and NM-Ag targets), due to the small crystal size and having similar lattice parameters across multiple phases. In the NM-Cu target XRD pattern, additional reflections from oxides (highlighted with question marks) can be determined, since copper is the element most prone to oxidation. As the intensity of these reflections is low in comparison to the reflection attributed to the metallic phases, we did not investigate them further. Oxidation was not observed for NM-Cu NPs, NM-Eq (target and NPs), and NM-Ag (target and NPs).

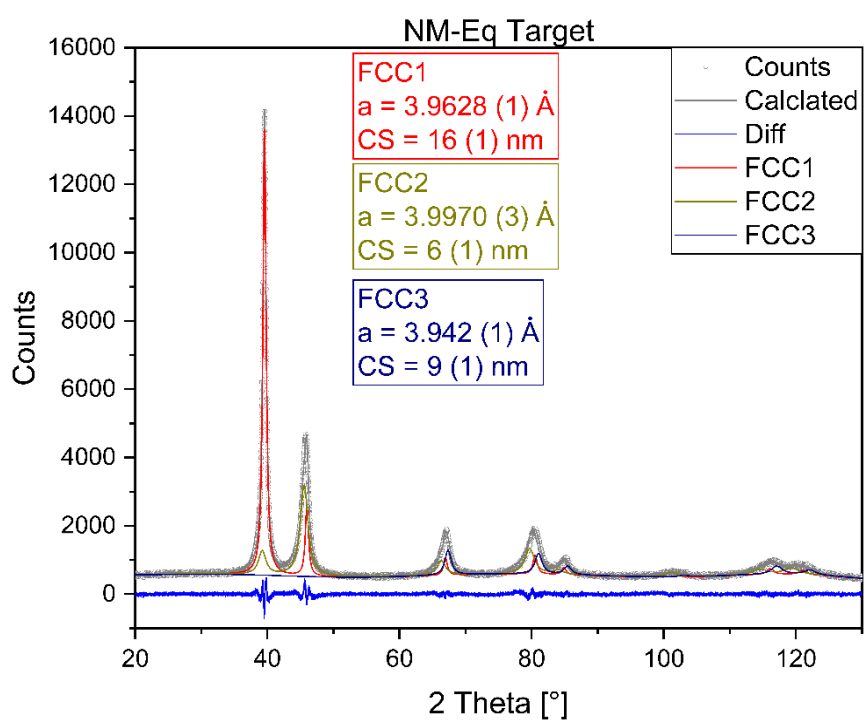

**Figure S10** Rietveld refinement for NM-Eq target.

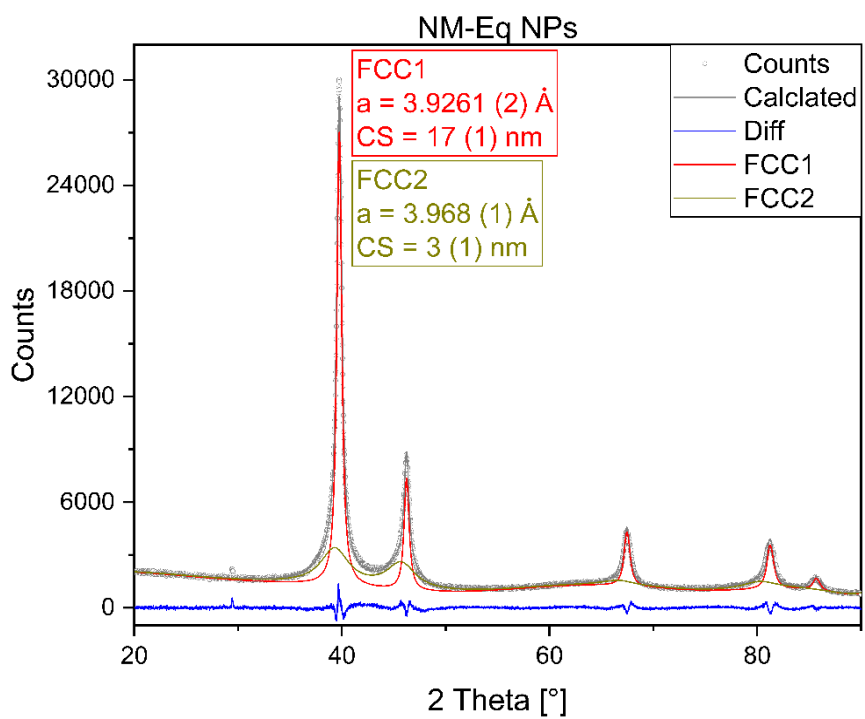

**Figure S11** Rietveld refinement for NM-Eq NP.

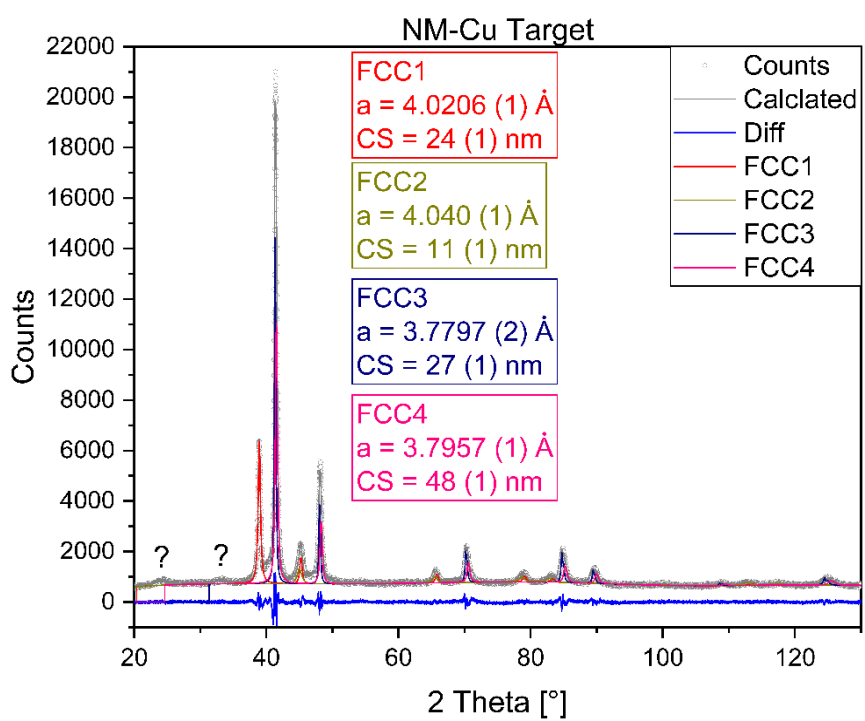

**Figure S12** Rietveld refinement for NM-Cu target, reflections emphasized with a question mark are oxides, which were not further determined, as their intensity in comparison to the reflection

attributed to the metallic phases is low. Due to the similar lattice parameters of FCC1 and FCC2 (FCC3 and FCC4), they are combined into two phases as described above.

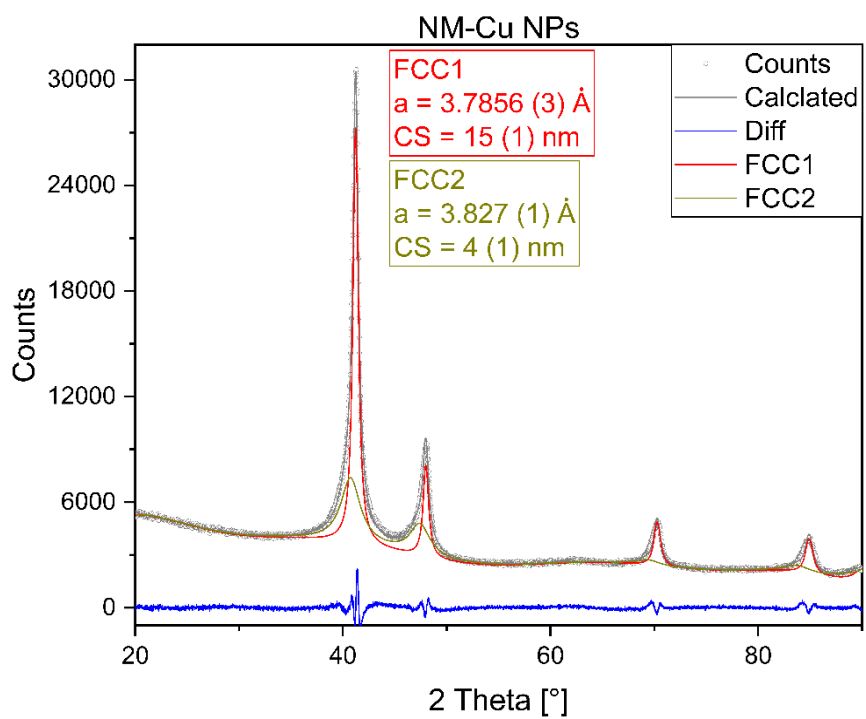

**Figure S13** Rietveld refinement for NM-Cu NP.

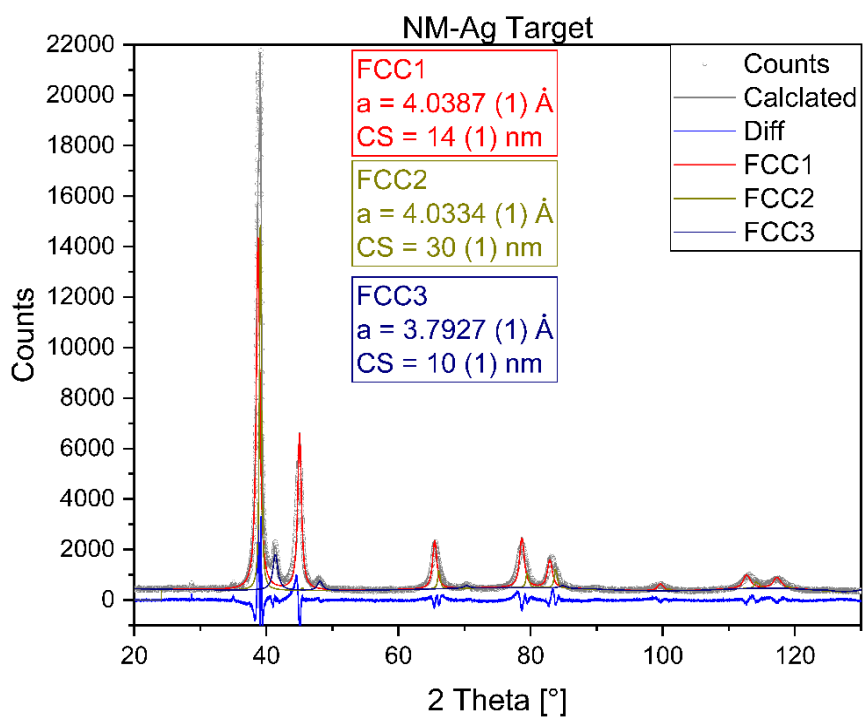

**Figure S14** Rietveld refinement for NM-Ag target. The (111) reflection shows a splitting, while the other reflections broaden and are asymmetric, resulting in the two similar FCC phases.

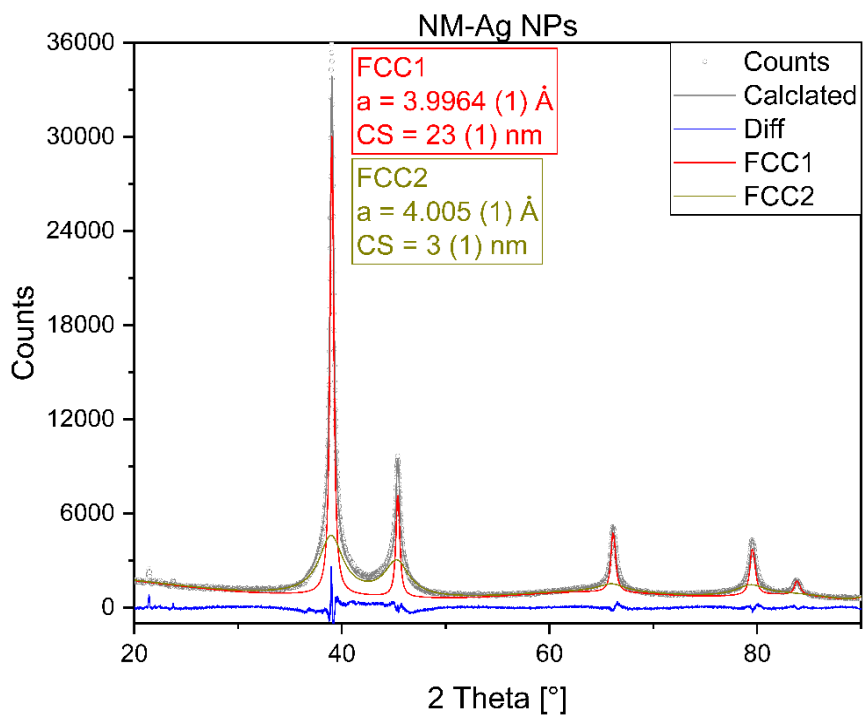

**Figure S15** Rietveld refinement for NM-Ag NP.

## Section S3: EDS large area measurements of nanoparticles and XPS spectra

**Table S2** EDS large area measurements of NM-Eq NPs

| Element | Cu [at-%] | Pd [at-%] | Ag [at-%] | Pt [at-%] | Au [at-%] |
|---|---|---|---|---|---|
| | 27.8 | 11.6 | 22.4 | 10.2 | 28.1 |
| | 31.0 | 11.9 | 21.8 | 7.8 | 27.4 |
| | 32.5 | 12.7 | 8.8 | 10.8 | 35.2 |
| | 29.5 | 10.5 | 22.7 | 7.2 | 30.0 |
| | 21.2 | 11.0 | 27.7 | 7.8 | 32.2 |
| | 29.2 | 12.1 | 23.1 | 8.3 | 27.4 |
| | 25.6 | 11.0 | 24.4 | 9.1 | 29.8 |
| **Mean** | 28.2 | 11.6 | 21.4 | 8.8 | 29.9 |
| **Standard Deviation** | 3.5 | 0.7 | 5.5 | 1.2 | 2.6 |

**Table S3** EDS large area measurements of NM-Cu NPs

| Element | Cu [at-%] | Pd [at-%] | Ag [at-%] | Pt [at-%] | Au [at-%] |
|---|---|---|---|---|---|
| | 57.5 | 9.8 | 12.1 | 7.9 | 12.8 |
| | 56.4 | 9.6 | 11.8 | 10.6 | 11.6 |
| | 55.1 | 10.1 | 15.1 | 7.5 | 12.2 |
| | 53.8 | 10.0 | 13.7 | 10.3 | 12.2 |
| | 57.0 | 10.1 | 13.7 | 8.8 | 10.4 |
| **Mean** | 56.0 | 9.9 | 13.3 | 9.0 | 11.8 |
| **Standard Deviation** | 1.3 | 0.2 | 1.2 | 1.2 | 0.8 |

**Table S4** EDS large area measurements of NM-Ag NPs

| Element | Cu [at-%] | Pd [at-%] | Ag [at-%] | Pt [at-%] | Au [at-%] |
|---|---|---|---|---|---|
| | 11.0 | 11.7 | 47.7 | 11.2 | 18.1 |
| | 10.8 | 11.5 | 48.4 | 10.0 | 19.4 |
| | 14.8 | 10.4 | 46.9 | 10.2 | 17.7 |
| | 16.3 | 9.1 | 45.5 | 9.8 | 17.7 |
| | 10.1 | 11.2 | 48.1 | 11.4 | 19.3 |
| | 14.2 | 11.3 | 47.9 | 8.1 | 18.5 |
| | 13.7 | 8.9 | 48.4 | 9.8 | 19.2 |
| | 11.0 | 11.9 | 50.6 | 8.8 | 17.7 |
| | 15.3 | 13.3 | 46.9 | 7.9 | 16.7 |
| **Mean** | 13.0 | 11.0 | 47.8 | 9.7 | 18.3 |
| **Standard Deviation** | 2.1 | 1.1 | 1.4 | 1.0 | 0.7 |

**Table S5** Calculated entropy of mixing $\Delta S_{mix}$ for NM-Eq, NM-Cu, and NM-Ag. The MC and MD data refer to simulated NPs.

|  | Nominal equimolar | NM-Eq | Nominal Cu-enriched | NM-Cu | Nominal Ag-enriched | NM-Ag |
|---|---|---|---|---|---|---|
| $\Delta S_{mix}$ | 1.6 $R$ | 1.5 $R$ | 1.4 $R$ | 1.3 $R$ | 1.4 $R$ | 1.4 $R$ |
| MC | 1.46 $R$ |  | 1.29 $R$ |  | 1.26 $R$ |  |
| MD | 1.48 $R$ |  | 1.34 $R$ |  | 1.29 $R$ |  |

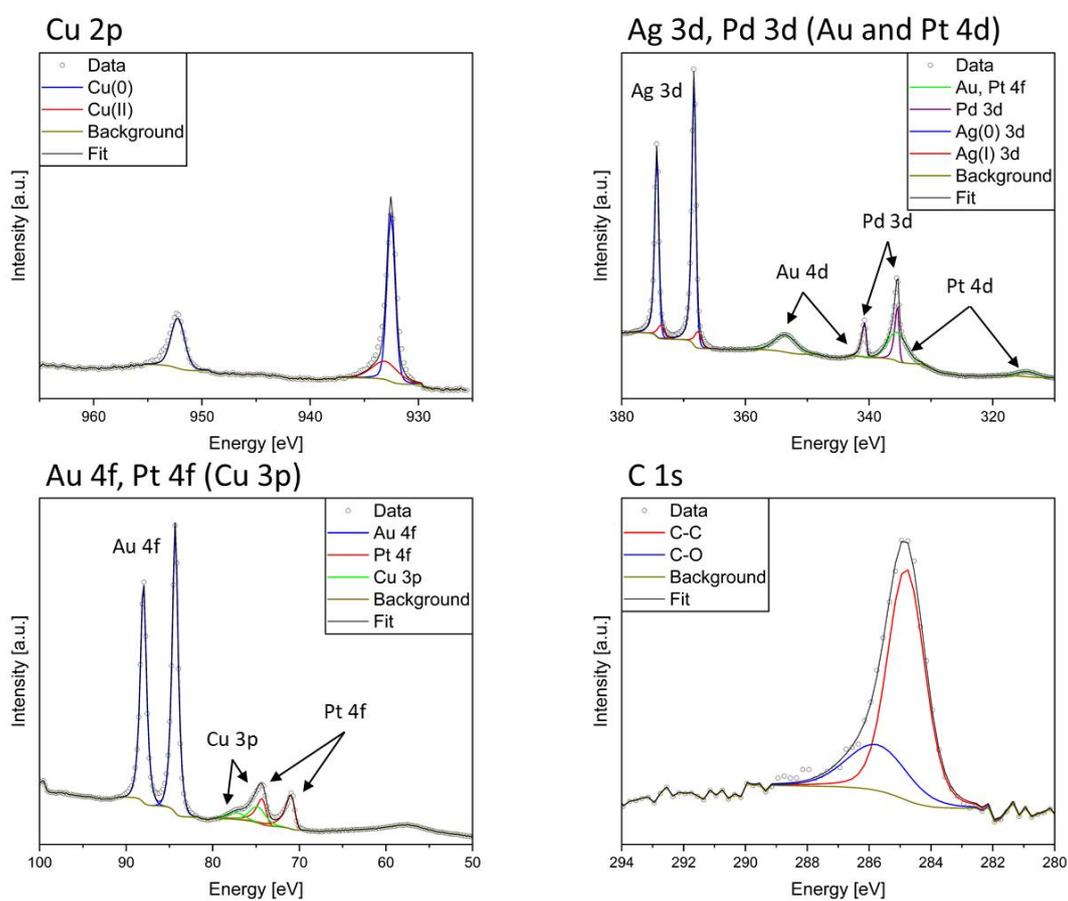

**Figure S16** Representative XPS high-resolution spectra of Cu 2p, Ag 3d, Pd 3d, Au 4f, Pt 4f, and C 1s corresponding to the dried colloid of NM-Eq. Note that the intensity scale is not equal in the diagrams, so peak intensity comparison between different diagrams is limited.

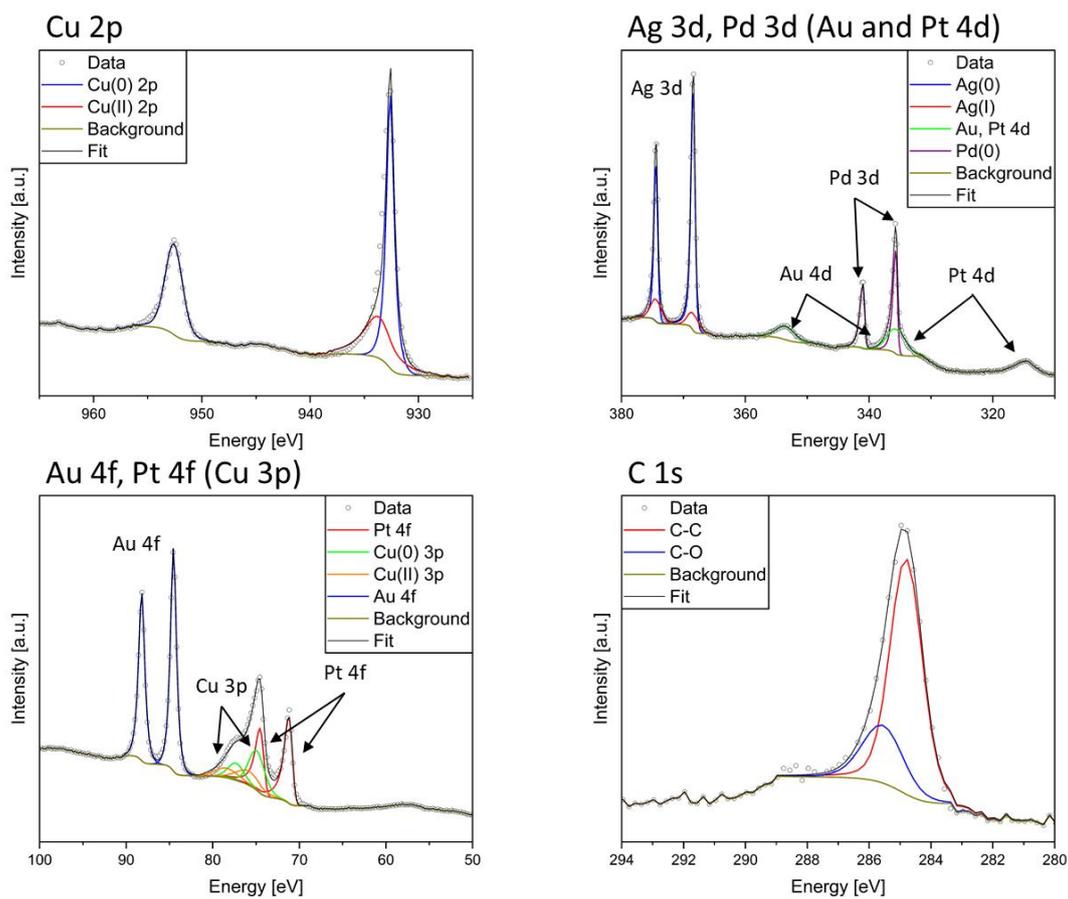

**Figure S17** Representative XPS high-resolution spectra of Cu 2p, Ag 3d, Pd 3d, Au 4f, Pt 4f, and C 1s corresponding to the dried colloid of NM-Cu. Note that the intensity scale is not equal in the diagrams, so peak intensity comparison between different diagrams is limited.

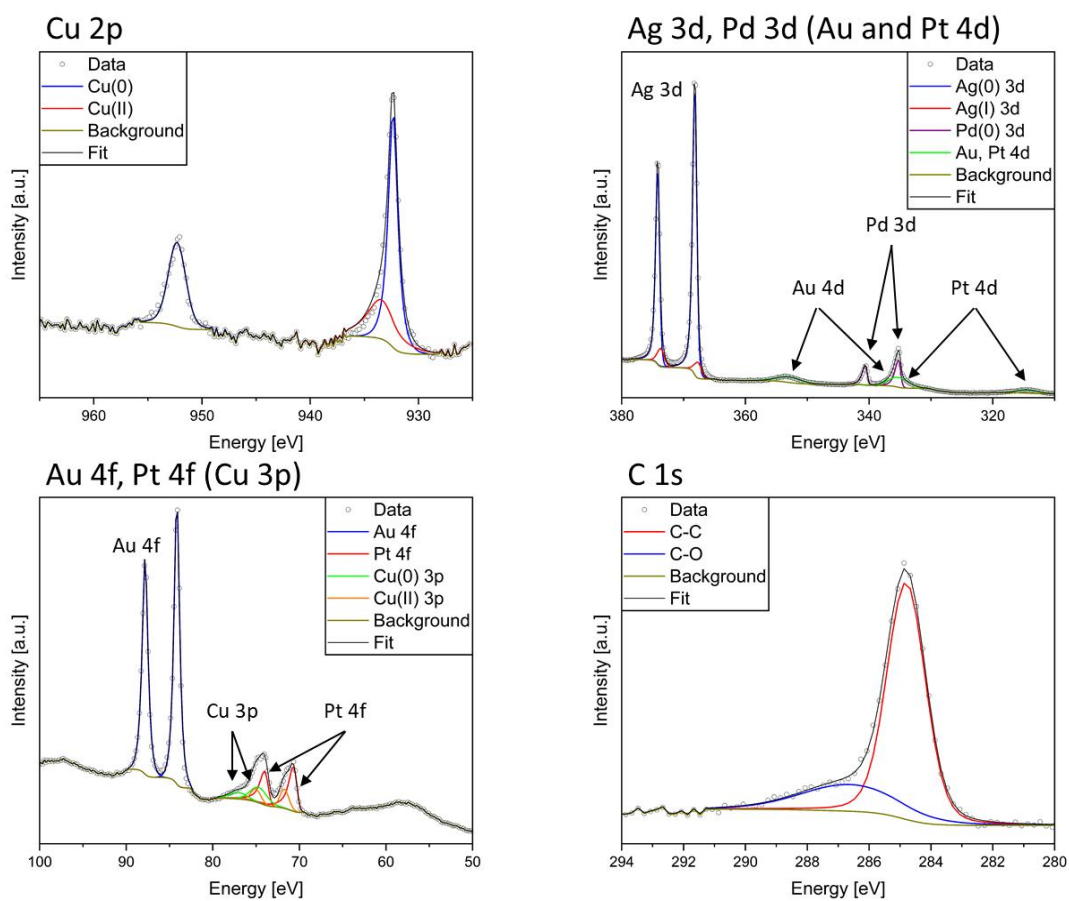

**Figure S18** Representative XPS high-resolution spectra of Cu 2p, Ag 3d, Pd 3d, Au 4f, Pt 4f, and C 1s corresponding to the dried colloid of NM-Ag. Note that the intensity scale is not equal in the diagrams, so peak intensity comparison between different diagrams is limited.

**Table S6** Determined composition from XPS for NM-Eq, NM-Ag, and NM-Cu

|       | Cu | Cu(II) | Pd | Ag | Ag(I) | Pt | Au |
|-------|----|--------|----|----|-------|----|----|
| **NM-Eq** | 32 | 4  | 8  | 31 | 2 | 4 | 19 |
| **NM-Ag** | 14 | 6  | 10 | 56 | 0 | 4 | 10 |
| **NM-Cu** | 38 | 28 | 7  | 14 | 1 | 5 | 7  |

**Consideration to theoretically possible diffusion**

The rapid cooling during particle formation can limit the diffusion. Considering a potential diffusion length of Ag atoms in Pt (solid) during this cooling phase, one can estimate whether segregation can theoretically occur inside our NPs or not. Assuming an effective cooling rate $\alpha$ of $10^{11}$ to $10^{13}$ Ks$^{-1}$ initially reached during ultra-short pulse irradiation in LAL and a diffusion constant $D$ of 0.13 cm$^2$s$^{-1}$ [1] for Ag in the temperature range 1473 to 1873 K ($\Delta T = 400$ K), the resulting cooling time $t = \Delta T/\alpha$ would be linearly approximated 0.04 to 4 ns, depending on the cooling rate (which depends on both the droplet size and its position in the plume, namely distance to the liquid´s vapor) [2],[3]. During this time Ag atoms could diffuse a distance $L = \sqrt{2 \cdot D \cdot t}$ of 32 to 320 nm. This would be sufficient for a phase segregation to occur.

## Section S4: Composition of individual nanoparticles

**Table S7** Composition of individual NM-Eq particles with marked outliers: high Ag content (green) and low Ag content (orange)

| Diameter [nm] | Cu [at-%] | Pd [at-%] | Ag [at-%] | Pt [at-%] | Au [at-%] |
|---|---|---|---|---|---|
| 131 | 34 | 14 | 21 | 6 | 25 |
| 117 | 31 | 17 | 22 | 7 | 23 |
| 82 | 29 | 13 | 32 | 7 | 19 |
| 62 | 33 | 15 | 22 | 8 | 23 |
| 59 | 31 | 12 | 27 | 8 | 23 |
| 57 | 27 | 11 | 36 | 6 | 21 |
| 53 | 35 | 17 | 19 | 6 | 23 |
| 53 | 31 | 12 | 29 | 7 | 21 |
| 52 | 29 | 14 | 30 | 6 | 21 |
| 46 | 29 | 13 | 28 | 6 | 24 |
| 45 | 29 | 13 | 31 | 7 | 20 |
| 43 | 26 | 11 | 39 | 5 | 18 |
| 40 | 32 | 14 | 24 | 9 | 22 |
| 38 | 30 | 13 | 27 | 7 | 22 |
| 36 | 36 | 16 | 12 | 9 | 27 |
| 35 | 22 | 15 | 29 | 7 | 27 |
| 33 | 27 | 17 | 16 | 11 | 30 |
| 31 | 25 | 17 | 20 | 11 | 27 |
| 30 | 26 | 13 | 34 | 6 | 22 |
| 29 | 26 | 14 | 34 | 6 | 19 |
| 24 | 24 | 14 | 34 | 8 | 21 |
| 23 | 33 | 13 | 22 | 4 | 28 |
| 23 | 23 | 15 | 33 | 11 | 19 |
| 22 | 15 | 17 | 38 | 7 | 23 |
| 21 | 27 | 12 | 36 | 6 | 20 |
| 17 | 18 | 14 | 52 | 3 | 14 |
| 16 | 21 | 14 | 36 | 6 | 23 |
| 15 | 13 | 17 | 43 | 5 | 23 |
| 14 | 13 | 20 | 34 | 9 | 25 |
| 11 | 3 | 9 | 85 | 4 | 1 |
| 10 | 13 | 19 | 40 | 8 | 20 |
| 9 | 16 | 12 | 57 | 2 | 1 |
| **Mean** | 25.2 | 14.2 | 32.6 | 6.8 | 21.1 |
| **Standard Deviation** | 7.6 | 2.5 | 13.3 | 2.1 | 6.1 |

**Table S8** Composition of individual NM-Cu particles with marked outliers: low Ag content (orange)

| Diameter [nm] | Cu [at-%] | Pd [at-%] | Ag [at-%] | Pt [at-%] | Au [at-%] |
|---|---|---|---|---|---|
| **121** | **54** | **18** | **2** | **15** | **11** |
| 117 | 50 | 15 | 17 | 8 | 12 |
| 97 | 52 | 15 | 13 | 9 | 11 |
| 96 | 51 | 16 | 11 | 9 | 14 |
| 87 | 50 | 14 | 19 | 7 | 11 |
| 77 | 54 | 15 | 12 | 12 | 7 |
| 77 | 53 | 14 | 14 | 10 | 9 |
| 66 | 48 | 15 | 16 | 8 | 12 |
| 65 | 50 | 15 | 16 | 8 | 11 |
| 64 | 52 | 15 | 16 | 7 | 11 |
| 51 | 50 | 15 | 13 | 11 | 11 |
| **49** | **53** | **18** | **3** | **17** | **9** |
| 48 | 48 | 15 | 16 | 9 | 12 |
| 45 | 52 | 16 | 8 | 16 | 8 |
| 42 | 50 | 16 | 12 | 8 | 15 |
| 41 | 48 | 15 | 18 | 8 | 10 |
| 40 | 57 | 16 | 11 | 8 | 9 |
| 40 | 50 | 15 | 17 | 9 | 10 |
| 38 | 50 | 17 | 16 | 7 | 10 |
| **37** | **53** | **17** | **4** | **17** | **9** |
| 36 | 55 | 14 | 10 | 9 | 13 |
| 34 | 55 | 17 | 9 | 10 | 10 |
| 34 | 48 | 18 | 12 | 10 | 12 |
| 33 | 48 | 16 | 18 | 7 | 11 |
| 31 | 49 | 13 | 19 | 7 | 12 |
| 25 | 50 | 17 | 8 | 9 | 15 |
| 20 | 53 | 14 | 14 | 8 | 11 |
| 19 | 52 | 13 | 12 | 13 | 11 |
| 13 | 48 | 16 | 16 | 11 | 10 |
| Mean | 51.1 | 15.4 | 12.7 | 9.8 | 10.9 |
| Standard Deviation | 2.5 | 1.5 | 4.6 | 2.9 | 1.8 |

**Table S9** Composition of individual NM-Ag particles with marked outliers: high Ag content (green) and low Ag content (orange)

| Diameter [nm] | Cu | Pd | Ag | Pt | Au |
|---|---|---|---|---|---|
| 103 | 9 | 13 | 64 | 5 | 9 |
| 88 | 9 | 13 | 63 | 5 | 10 |
| 66 | 9 | 14 | 60 | 6 | 11 |
| 63 | 9 | 13 | 62 | 6 | 10 |
| 60 | 12 | 12 | 60 | 4 | 12 |
| 56 | 10 | 14 | 59 | 6 | 11 |
| 52 | 12 | 11 | 61 | 5 | 11 |
| 47 | 9 | 13 | 63 | 5 | 10 |
| 46 | 9 | 11 | 65 | 5 | 10 |
| 41 | 11 | 13 | 59 | 6 | 11 |
| 39 | 10 | 13 | 61 | 5 | 10 |
| 37 | 9 | 13 | 64 | 5 | 9 |
| 35 | 14 | 14 | 52 | 5 | 14 |
| 31 | 9 | 16 | 54 | 9 | 12 |
| 31 | 8 | 12 | 67 | 5 | 8 |
| 31 | 11 | 19 | 45 | 12 | 14 |
| 31 | 9 | 12 | 64 | 5 | 10 |
| 30 | 11 | 10 | 66 | 4 | 10 |
| 30 | 9 | 10 | 66 | 5 | 9 |
| 29 | 11 | 12 | 63 | 5 | 9 |
| 29 | 11 | 14 | 57 | 7 | 12 |
| 28 | 8 | 14 | 62 | 6 | 10 |
| 25 | 5 | 10 | 77 | 2 | 5 |
| 24 | 8 | 13 | 63 | 6 | 10 |
| 23 | 15 | 25 | 26 | 14 | 21 |
| 23 | 7 | 12 | 65 | 6 | 9 |
| 22 | 5 | 10 | 74 | 4 | 6 |
| 22 | 7 | 10 | 68 | 5 | 10 |
| 21 | 10 | 12 | 64 | 6 | 9 |
| 21 | 8 | 11 | 70 | 4 | 8 |
| 20 | 10 | 27 | 29 | 16 | 18 |
| 19 | 7 | 12 | 71 | 5 | 6 |
| 18 | 4 | 16 | 71 | 3 | 5 |
| 17 | 8 | 15 | 61 | 6 | 9 |
| 16 | 9 | 14 | 62 | 6 | 9 |
| 16 | 13 | 18 | 53 | 9 | 7 |
| 16 | 5 | 24 | 59 | 6 | 5 |
| 16 | 9 | 11 | 73 | 3 | 4 |
| 15 | 11 | 31 | 20 | 14 | 26 |
| 14 | 12 | 17 | 53 | 7 | 12 |
| 13 | 5 | 5 | 86 | 1 | 2 |
| 12 | 9 | 11 | 73 | 3 | 5 |
| 11 | 5 | 25 | 43 | 19 | 9 |
| Mean | 9.2 | 14.3 | 60.3 | 6.3 | 9.9 |
| Standard Deviation | 2.5 | 4.9 | 12.4 | 3.5 | 4.2 |

## Section S5: EDS-elemental mapping of targets

EDS elemental maps of target surfaces for NM-Eq (Figure S2), NM-Cu (Figure S3), and NM-Ag (Figure S4)

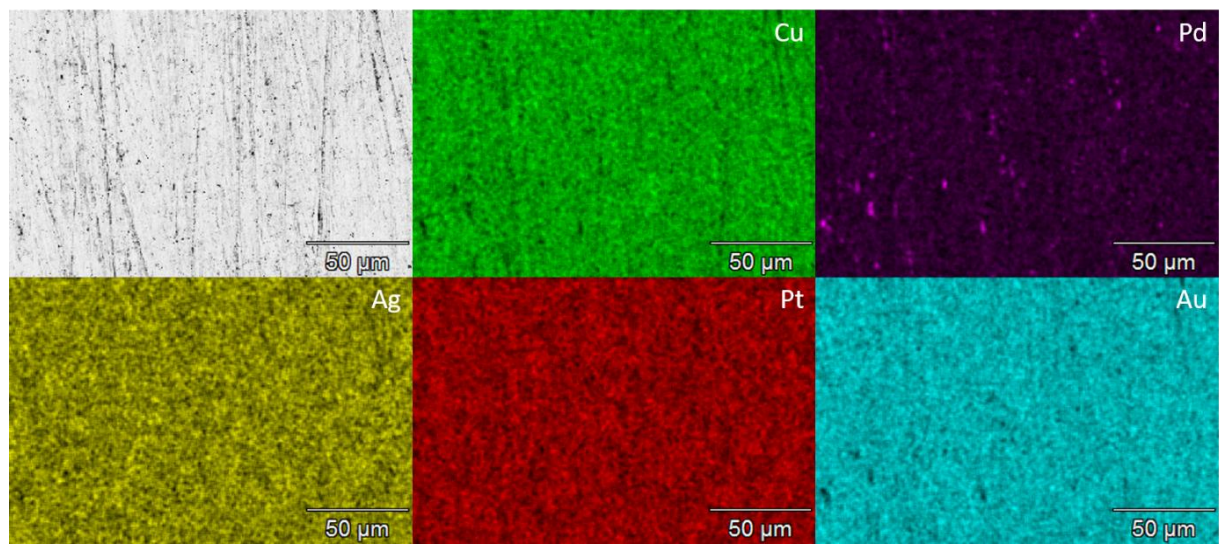

**Figure S19** EDS elemental map of NM-Eq target.

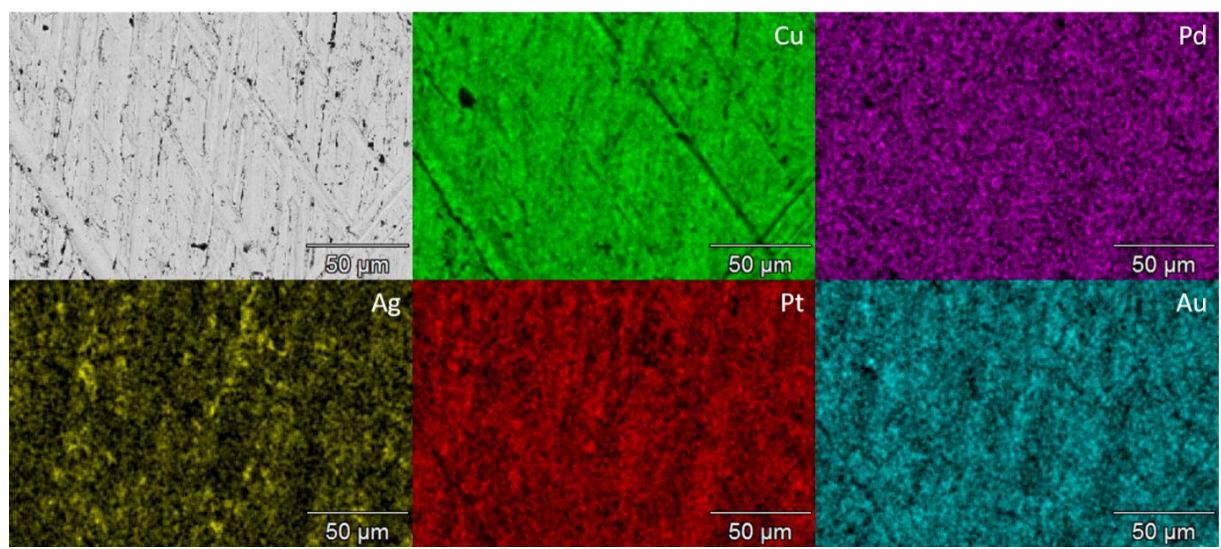

**Figure S20** EDS elemental map of NM-Cu target.

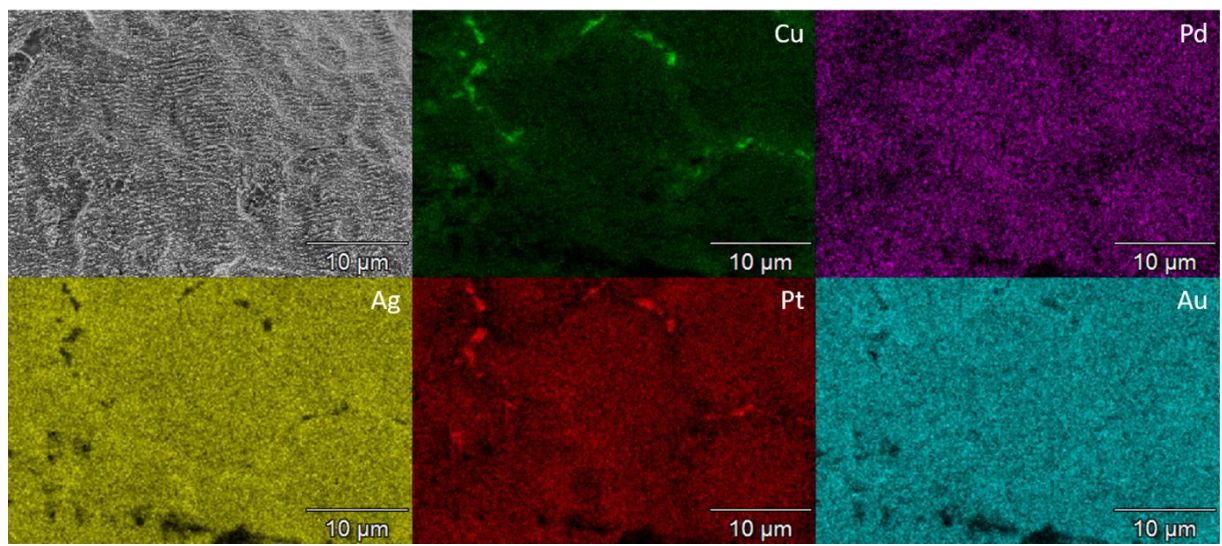

**Figure S21** EDS elemental map of NM-Ag target.

**Section S6: Selected SAED patterns from *in situ* heating measurement of NM-Cu**

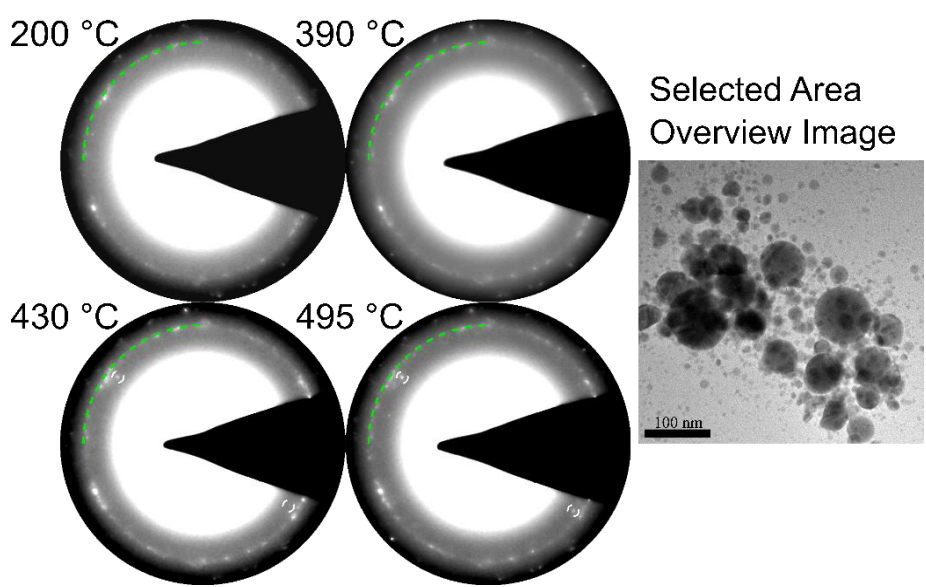

**Figure S22** Selected SAED patterns (edited contrast/brightness due to low intensity from reflections) from *in situ* heating measurements of several NPs shown post-heating in the selected area overview image: the green dashed line marks reflections present pre-heating, the white circle marks additional reflections upon heating. Overview image on the right showing NPs in the selected area for electron diffraction.

**Supporting Information References**

[1] E. A. Brandes and G. B. Brook, Eds., Smithells metals reference book, 7th ed. / edited by E.A. Brande and G.B. Brook. Oxford: Butterworth-Heinemann, 1999.